\documentstyle[12pt]{article}

\title{METHODS OF MINIMIZATION OF CALCULATIONS IN HIGH ENERGY
       PHYSICS: \\
       {\sf I.}~A Covariant Method of the Calculation of
Amplitudes of Processes Involving Polarized Dirac Particles}

\author {Alexander~L.~Bondarev
\and \it National Scientific and Educational Center of Particle and
\and \it High Energy Physics attached to Belarusian State University
\and \it M.Bogdanovich str.,153, Minsk 220040, Republic of Belarus
\and \rm e{\it-mail}: bondarev@hep.by}

\begin{document}
\maketitle

\begin{abstract}

    The article deals  with a number  of the existing  variants of
direct calculation of amplitudes of processes with polarized Dirac
particles.  It is shown, that all of them are special cases of one
and   the   same   mathematical   scheme.     The  advantages  and
disadvantages of this scheme are considered.  A new variant of the
method of calculation of amplitudes, keeping all the advantages of
this  scheme,  but  free  from  disadvantages,  is  proposed.   In
particular,  this  variant  is  suitable  for  the  evaluation  of
amplitudes of  processes with  interfering diagrams  and does  not
need  any  additional  calculations,  that  are  necessary  in the
general  case.    Expressions  for  the  amplitudes  of  processes
involving  both  massive  and  massless  particles  are presented.
These expressions make  it possible to  create in an  easy way the
computer programs  for automatic  calculation of  amplitudes.  The
existing computer  programs for  calculation of  amplitudes, their
limitations and disadvantages are briefly considered.

\end{abstract}


\section {Introduction}

    In order to  calculate the observables  of the processes  with
the Dirac particles, especially for high-order diagrams and in the
case  of  taking  into  account  the  polarization  effects, it is
necessary to calculate the traces  of products of great number  of
the  Dirac  $\gamma$-matrices.    Thus,  we  face  the  problem of
obtaining the analytical  expressions for the  physical quantities
we are interested in.

    One  way  to  eliminate  this  problem  is  to  calculate  the
amplitudes of the  processes directly.   This method allows  as to
reduce essentially  the number  of the  Dirac $\gamma$-matrices in
the  expressions considered so also the number  of the expressions
themselves  (in  the  situation  when  we  have  the   interfering
diagrams).    The  results  obtained  in  this  way depend only on
4-vectors and are suitable  for numerical analysis.   To calculate
the  squared  amplitude  (by   means  of  which  observables   are
expressed), we need to calculate  only the squares of the  modules
of the complex numbers obtained.  The numerical values obtained in
this way are equivalent in principle to the values obtained in the
classical  way,  but  can  have  smaller  errors caused by loss of
accuracy at the calculations.

    The  appeal  of  such  an  approach  has  stimulated  a lot of
articles,  where  a  number  of  different  ways of calculation of
amplitudes  have  been  proposed.    However  all of them have the
limitations.    Moreover,  there  are  the  incorrect methods (for
example  \cite{rm.1-1}~--~\cite{rm.1-3},   see  subsection   6  of
Section~{\bf 6}).  These reasons and the fact that the expressions
which have been obtained by different ways for the same amplitudes
have different analytical form lead  to a certain mistrust to  the
idea of direct calculation of amplitudes.

    Up to now the methods  of calculation of amplitudes have  been
separated  into  two  classes:    the  covariant methods, i.e. the
methods reduced to the evaluation of the traces of the products of
the Dirac  $\gamma$-matrices (see  \cite{b1}~--~\cite{r2.10b}) and
the methods reduced to the multiplication of $\gamma$-matrices and
bispinors   which   are   written   in   the   matrix   form  (see
\cite{r6.1.2-3}~--~\cite{r6.1.4-1},
\cite{r6.1.4-3}~--~\cite{r6.2-2}).    However  all  of them are in
fact the special cases of one and the same mathematical scheme.

    In Section~{\bf  2} the  covariant methods  of calculation are
briefly   considered.      The   general  mathematical  scheme  of
calculation  of  amplitudes  is  presented.   It describes all the
known methods.  It  is shown also how  the results of any  special
method can be transformed into the results of another one.

    In Section~{\bf  3} the  advantages and  disadvantages of  the
general  scheme  are  considered.    In  particular  we   consider
carefully the  additional calculations  which are  necessary to be
performed  in  the  general  case  when  one  has  the interfering
diagrams.

    In Section~{\bf 4} a new  covariant  method of  calculation of
amplitudes is proposed.  This method  keeps all  the advantages of
the general scheme but is free from its disadvantages.

    In Section~{\bf 5} the general expressions for the  amplitudes
of processes involving both  massive and massless Dirac  particles
are presented.  (Examples of calculation of the concrete processes
can be found in \cite{b1}.)  Using the presented formulae one  can
create  the   computer  program   for  automatic   calculation  of
amplitudes.

    In Section~{\bf  6} the  methods of  calculation of amplitudes
based  on  the  multiplication  of $\gamma$-matrices and bispinors
which are written in the matrix form and some similar methods  are
considered.  It is shown that all of them are the special cases of
the general scheme considered.

    In  Section~{\bf  7}   the  existing  computer   programs  for
automatic  calculation  of   amplitudes,  their  limitations   and
drawbacks are briefly considered.


\section {The general scheme of the covariant calculation of
amplitudes}

    There is  an even  number $(2N)$  of fermions  in initial  and
final  state  for  any  reaction  with Dirac particles.  Therefore
every  diagram  contains  $N$   nonclosed  fermion  lines.     The
expression
\begin {equation}
\displaystyle
 M_{if} = \bar{u}_f Q u_i
\label{e2.1}
\end {equation}
    corresponds to  every line  in the  amplitude of  the process,
$
\displaystyle
\,
u_i
$,
$
\displaystyle
u_f
$
    are   the   Dirac   bispinors   for   free  particles.    (For
definiteness, we assume that fermions are particles.  However  the
results obtained can be generalized  to the case when both  of the
fermions  are  antiparticles  or  one  fermion  is  a particle and
another one is an antiparticle.  This generalization is considered
at the end of Section~{\bf 4}.)
$$
\displaystyle
\bar{u} = u^{+} \gamma^0
\, ,
$$
    and $Q$ is a  matrix operator characterizing the  interaction.
The  operator  $Q$  is  expressed  as  a linear combination of the
products  of  the  Dirac  $\gamma$-matrices (or their contractions
with 4-vectors) and can have any number of free Lorentz indexes.

    The formula (\ref{e2.1}) can be rewritten as
$$
\displaystyle
M_{if} = Tr ( Q u_i \bar{u}_f )
$$
    however the expression for the operator
$
\displaystyle
u_i \bar{u}_f
$
    is not known.  Because of this, in order to calculate $M_{if}$
we use the following general scheme
\begin {equation}
\begin {array}{l} \displaystyle
M_{if} = \bar{u}_f Q u_i = ( \bar{u}_f Q u_i ) \cdot
       { \bar{u}_i Z u_f \over \bar{u}_i Z u_f }
   = { Tr ( Q u_i \bar{u}_i Z u_f \bar{u}_f ) \over
                    \bar{u}_i Z u_f }
                        \\[0.5cm] \displaystyle
\simeq { Tr ( Q u_i \bar{u}_i Z u_f \bar{u}_f ) \over
         | \bar{u}_i Z u_f | }
     = { Tr ( Q u_i \bar{u}_i Z u_f \bar{u}_f ) \over
\sqrt{ Tr ( \bar{Z} u_i \bar{u}_i Z u_f \bar{u}_f ) } }
   = {\cal M}_{if}
\end {array}
\label{e2.2}
\end {equation}
    where $Z$ is an arbitrary $4 \times 4$-matrix,
$$
\displaystyle
\bar{Z} = \gamma^0 Z^{+} \gamma^0
$$
    (symbol $\simeq $ stands for "an equality up to a phase factor
sign").

    The  projection  operators  are  substituted for $u\bar{u}$ in
(\ref{e2.2}).  For a particle with mass $m$ we have:
\begin {equation}
\displaystyle
u(p,n) \bar{u}(p,n) = { 1 \over 2 } ( \hat{p} + m )
                      ( 1 + \gamma_5 \hat{n} ) = {\cal P}(p,n)
\label{e2.3}
\end {equation}
    where%
\footnote{
We use the same metric as in the book \cite{r6.1.4-2}:
$$
\displaystyle
a^{\mu} = ( a_0 , \vec{a} ) ,
\;\;\;
a_{\mu} = ( a_0, -\vec{a} ) ,
\;\;\;
ab = a_{\mu} b^{\mu} = a_0 b_0 - \vec{a} \vec{b}
$$
}:
$
\displaystyle
\hat{a} = \gamma_{\mu} a^{\mu}
\;\;
$
    for any 4-vector $a$,
$$
\displaystyle
p^2 = m^2
\; , \;\;
n^2 = -1
\; , \;\;
pn = 0
\; , \;\;
\bar{u} u = 2 m
\; , \;\;\;
\gamma_5 = {\it i} \gamma^0 \gamma^1 \gamma^2 \gamma^3
$$
    ($p$ is the  4-momentum of the  particle, $n$ is  the 4-vector
specifying the axis of the spin projections of the particle).

    For  a  massless  particle  the  projection  operator  is  the
following
\begin {equation}
\displaystyle
u_{\pm}(p) \bar{u}_{\pm}(p)
= { 1 \over 2 }( 1 \pm \gamma_5 ) \hat{p} = {\cal P}_{\pm}(p)
\label{e2.4}
\end {equation}
    where
$$
\displaystyle
p^2 = 0
\; , \;\;
\bar{u}_{\pm} \gamma_{\mu} u_{\pm} =  2 p_{\mu}
$$
    (signs $\pm $ correspond to the helicity of particle) .

    Actually the general scheme (\ref{e2.2}) describes all
the known methods of  calculation of amplitudes.
\begin{description}
\item[1.]
    In the articles \cite{r2.1-1}~--~\cite{r2.1-5c} one chooses
$$
\displaystyle
Z = 1
\; .
$$
    The results of the articles \cite{r2.1-6} reduce to the same
choice for the calculation of amplitudes of type
$
\displaystyle
\bar{u}_{\pm}(p_f) u_{\mp}(p_i)
$
    for the processes with massless particles.
\item[2.]
    In \cite{r2.1-5b}, \cite{r2.1-5c} one proposes
$$
\displaystyle
Z = \gamma_5
\; .
$$
    The results of  the article \cite{r2.2-1}  reduce to the  same
choice.
\item[3.]
    The results of \cite{r2.3-1}, \cite{r2.3-2} reduce to
$$
\displaystyle
Z = \gamma^0
\;\; .
$$
    The expressions obtained in the article \cite{r2.3-3}
reduce to the same choice under the following restrictions:
\begin{description}
\item[(a)]
    Dirac particles  with equal  mass in  the center-of-mass frame
are considered;
\item[(b)]
    the polarization state of the particles is the helicity,  i.e.
4-vectors, which determine the axes of spin projections, are  (see
e.g.  \cite{g1}):
$$
\displaystyle
n^{\mu}(p) = { \lambda \over m } \left( \left| \vec{p} \right| ,
  { p_0 \over \left| \vec{p} \right| } \vec{p} \right)
\;\; , \;\;\;
\lambda = \pm 1
\; ;
$$
\item[(c)]
    the particles have the opposite helicities.
\end{description}
\item[4.]
    The results of \cite{r2.3-3} also reduce to the choice
$$
\displaystyle
Z = \gamma^0 \gamma_5 \hat{l}
$$
    [where
$  \,
\displaystyle
l^{\mu} = { 1 \over |\vec{p}| \sqrt{ {p_x}^2 + {p_y}^2 } }
 \left[ 0 , \, p_x p_z , \, p_y p_z , - ( {p_x}^2 + {p_y}^2 )
\right]
 , \,
l^2 = -1
 , \,
( l p_i ) = ( l p_f ) = 0
\, ]
$
\newline
    under the following restrictions:
\begin{description}
\item[(a)]
    Dirac particles  with equal  mass in  the center-of-mass frame
are considered;
\item[(b)]
    the polarization state of the particles is the helicity;
\item[(c)]
    the particles have equal helicities.
\end{description}
\item[5.1.]
    In \cite{r2.5.1-1a} it was proposed to choose for $Z$
$$
\displaystyle
Z = \hat{l}
$$
    where $l$ is an arbitrary 4-vector  such that
$
\displaystyle
l^2 = -1
$.
    The  same  choice  was  proposed  in  \cite{r2.1-4a}  for  the
calculation of amplitudes of processes with massless particles.
\item[5.2.]
    In \cite{r2.5.2-1} it was proposed to choose for $Z$
$$
\displaystyle
Z = \hat{q}
$$
    (where $q$ is an arbitrary 4-vector such that
$
\displaystyle
q^2 = 0
$,
    it was proposed to choose
$
\displaystyle
q^{\mu} = (1,1,0,0)
$
    for numerical calculations) under the following restrictions:
\begin{description}
\item[(a)]
    the  4-vectors,   which  determine   the  axes   of  the  spin
projections, are
$$
\displaystyle
n = \lambda \left[ { 1 \over m } p
                     - { m \over (pq) } q \right]
\;\; , \;\;
\lambda = \pm 1
$$
    ($p$ is the 4-momentum of the corresponding particle),
\item[(b)]
$$
\displaystyle
\lambda_i = \lambda_f
$$
\end{description}
\item[5.3.]
    In  \cite{r2.1-5c}   it  was   also  proposed   that  for  the
calculation of amplitudes of processes with massless particles
$$
\displaystyle
Z = \hat{r}
$$
    where $r$ is an arbitrary 4-momentum such that $ r^2 = m^2$.
\item[6.]
    The results of papers \cite{r2.6-1} -- \cite{r2.6-3} reduce to
$$
\displaystyle
Z = 1 + \gamma^0
\;\; .
$$
    The results of the paper \cite{r2.6-4} reduce to the same
choice under the restriction
$$
\displaystyle
n_f = \left[ ( n_f )_0 \: , \;\;
\vec{n}_i - { ( n_i )_0 \over ( p_i )_0 + m_i } \vec{p}_i
   + { ( n_f )_0 \over ( p_f )_0 + m_f } \vec{p}_f \right]
\;\; ,
$$
$$
\displaystyle
( n_f )_0 =  { 1 \over m_f } \left[
 ( \vec{p}_f \vec{n}_i )
- { ( n_i )_0 \over ( p_i )_0 + m_i }
 ( \vec{p}_f \vec{p}_i )  \right]
\;\; .
$$
\item[7.]
    The results of \cite{r2.6-4} also reduce to
$$
\displaystyle
Z = ( 1 + \gamma^0 ) \gamma_5 \hat{l}
$$
    where
$$
\displaystyle
n_f = \left[ - ( n_f )_0 \: , \;\;
- \vec{n}_i + { ( n_i )_0 \over ( p_i )_0 + m_i }
\vec{p}_i
   - { ( n_f )_0 \over ( p_f )_0 + m_f } \vec{p}_f \right]
\;\; ,
$$
$$
\displaystyle
( n_f )_0 = { 1 \over m_f } \left[
 ( \vec{p}_f \vec{n}_i )
- { ( n_i )_0 \over ( p_i )_0 + m_i }
 ( \vec{p}_f \vec{p}_i )  \right]
\;\; .
$$
    where
$$
\displaystyle
l^{\mu} = ( 0 , - \vec{o} )
\; , \; \;
{ \vec{o} \, }^2 = 1
\; , \; \;
( \vec{o} \vec{n}_i )
= { (n_i)_0 \over (p_i)_0 + m_i } ( \vec{o} \vec{p}_i )
\; .
$$
\item[8.]
    The results of papers \cite{r2.8-1}, \cite{r2.8-2} reduce to
$$
\displaystyle
Z = m + \hat{r}
$$
    ($r$ is an arbitrary 4-momentum such that
$
\displaystyle
r^2 = m^2
$),
    as this  takes place  in this  paper for  the 4-vectors, which
determine the axes of the spin projections, one uses
$$
\displaystyle
n_i = { m_i^2 p_f - ( p_i p_f ) p_i \over
   m_i \sqrt{ ( p_i p_f )^2 - m_i^2 m_f^2 } }
\;\;\; , \;\;\;\;
n_f = - { m_f^2 p_i - ( p_i p_f ) p_f \over
   m_f \sqrt{ ( p_i p_f )^2 - m_i^2 m_f^2 } }
\;\;.
$$
\item[9.]
    The results of \cite{r2.9} for the amplitudes of type
$
\displaystyle
\bar{u}_{\pm}(p_f) u_{\mp}(p_i)
$
    with massless particles reduce to
$$
\displaystyle
Z = \hat{q_1} \hat{q_2}
$$
    where $q_1$, $q_2$ are arbitrary 4-vectors  such that
$
\displaystyle
q_1^2 = q_2^2 = 0
$.
\item[10.]
    The results of the papers \cite{r2.10a}, \cite{r2.10b} for
the amplitudes of type
$
\displaystyle
\bar{u}_{\pm}(p_f) u_{\mp}(p_i)
$
    with massless particles reduce to
$$
\displaystyle
Z= \hat{l} \hat{q}
$$
    where $l$, $q$ are arbitrary 4-vectors  such that
$$
\displaystyle
l^2 = -1  \; ,\;\;  q^2= 0  \; , \;\; (ql)=0 \; .
$$
    As this takes place one chooses
$
\displaystyle
l^{\mu} = (0,0,1,0) , \;\; q^{\mu} = (1,1,0,0)
$
    for numerical calculations.

    The results of the paper \cite{r2.5.2-1} reduce to the same
choice for the calculation of amplitudes of processes with massive
particles under the following restrictions:
\begin{description}
\item[(a)]
    for  the  4-vectors,  which  determine  the  axes  of the spin
projections, one uses
$$
\displaystyle
n = \lambda \left[ { 1 \over m } p
                 - { m \over (pq) } q \right]
\;\; , \;\;
\lambda = \pm 1
$$
    ($p$ is the 4-momentum of the corresponding particle),
\item[(b)]
$$
\displaystyle
\lambda_i = - \lambda_f
$$
\end{description}
%
\end{description}
    The remaining  methods of  the calculation  of amplitudes  are
considered in Section~{\bf 6}.   We emphasize again that all  this
methods can be reduced to (\ref{e2.2}).

    The results of the  calculation of amplitudes for  different Z
differ only by the phase factor.  Really,
\begin {equation}
\begin {array}{l} \displaystyle
{\cal M}_{if} ( Z_1 )
    = { Tr ( Q {\cal P}_i Z_1 {\cal P}_f ) \over
\sqrt { Tr ( \bar{Z}_1 {\cal P}_i  Z_1 {\cal P}_f ) } }
    = { Tr ( Q {\cal P}_i Z_1 {\cal P}_f ) \over
\sqrt { Tr ( \bar{Z}_1 {\cal P}_i  Z_1 {\cal P}_f ) } }
\cdot
    { Tr ( \bar{Z}_2 {\cal P}_i Z_2 {\cal P}_f ) \over
      Tr ( \bar{Z}_2 {\cal P}_i Z_2 {\cal P}_f ) }
   \\       \\        \displaystyle
  = { Tr ( Q {\cal P}_i Z_1 {\cal P}_f ) \cdot
      Tr ( \bar{Z}_2 {\cal P}_i Z_2 {\cal P}_f )
   \over
\sqrt { Tr ( \bar{Z}_1 {\cal P}_i Z_1 {\cal P}_f ) \cdot
        Tr ( \bar{Z}_2 {\cal P}_i Z_2 {\cal P}_f ) }
\sqrt { Tr ( \bar{Z}_2 {\cal P}_i Z_2 {\cal P}_f ) } }
    \\      \\        \displaystyle
  = { Tr ( Q {\cal P}_i Z_2 {\cal P}_f ) \cdot
      Tr ( \bar{Z}_2 {\cal P}_i Z_1 {\cal P}_f )
   \over
\sqrt { Tr ( \bar{Z}_1 {\cal P}_i Z_2 {\cal P}_f ) \cdot
        Tr ( \bar{Z}_2 {\cal P}_i Z_1 {\cal P}_f ) }
\sqrt { Tr ( \bar{Z}_2 {\cal P}_i Z_2 {\cal P}_f ) } }
    \\      \\        \displaystyle
  = { Tr ( Q {\cal P}_i Z_2 {\cal P}_f ) \over
\sqrt { Tr ( \bar{Z}_2 {\cal P}_i  Z_2 {\cal P}_f ) } }
   \cdot   \left[
  { Tr ( \bar{Z}_2 {\cal P}_i Z_1 {\cal P}_f ) \over
    Tr ( \bar{Z}_1 {\cal P}_i Z_2 {\cal P}_f ) }
                  \right]^{1/2}
  = {\cal M}_{if}(Z_2) \cdot e^{ {\it i} \phi }
\;\; ,
\end {array}
\label{e2.5}
\end {equation}
    where
\begin {equation}
\displaystyle
e^{ {\it i} \phi }
=  \left[
  { Tr ( \bar{Z}_2 {\cal P}_i Z_1 {\cal P}_f ) \over
    Tr ( \bar{Z}_1 {\cal P}_i Z_2 {\cal P}_f ) }
                  \right]^{1/2}
\;\; .
\label{e2.6}
\end {equation}
    In (\ref{e2.5}) we have used the identity:
\begin {equation}
\begin {array}{c} \displaystyle
Tr ( A {\cal P}_i B {\cal P}_f ) \cdot
Tr ( C {\cal P}_i D {\cal P}_f )
= Tr ( A u_i \bar{u}_i B u_f \bar{u}_f ) \cdot
  Tr ( C u_i \bar{u}_i D u_f \bar{u}_f )
                        \\[0.5cm] \displaystyle
= ( \bar{u}_f A u_i ) ( \bar{u}_i B u_f )
  ( \bar{u}_f C u_i ) ( \bar{u}_i D u_f )
\equiv ( \bar{u}_f A u_i ) ( \bar{u}_i D u_f )
       ( \bar{u}_f C u_i ) ( \bar{u}_i B u_f )
                        \\[0.5cm] \displaystyle
= Tr ( A u_i \bar{u}_i D u_f \bar{u}_f ) \cdot
  Tr ( C u_i \bar{u}_i B u_f \bar{u}_f )
= Tr ( A {\cal P}_i D {\cal P}_f ) \cdot
  Tr ( C {\cal P}_i B {\cal P}_f )
\end {array}
\label{e2.7}
\end {equation}
    where  $A$,  $B$,  $C$,  $D$  are  arbitrary  $  4  \times   4
$-matrices.


\section {Advantages and disadvantages of the general scheme}

    Let  us  consider  the  advantages  and  disadvantages  of the
general   scheme   of   calculation   of  amplitudes  (\ref{e2.2})
illustrating them by the elementary example with
$
\displaystyle
 Z = 1
\;\; .
$

    The advantages are following:
\begin{description}
\item[1.]
    The calculation of the  amplitude of the process  is analogous
to the calculation of the squared amplitude of this process
\begin {equation}
\displaystyle
|M_{if}|^2
= ( \bar{u}_f Q u_i ) ( \bar{u}_f Q u_i )^{*}
= ( \bar{u}_f Q u_i ) ( \bar{u}_i \bar{Q} u_f )
= Tr ( Q u_i \bar{u}_i \bar{Q} u_f \bar{u}_f )
= Tr ( Q {\cal P}_i \bar{Q} {\cal P}_f )
\label{e3.1}
\end {equation}
    and reduces to the standard operation of the evaluation of the
trace  of  a  linear  combination  of  the  products  of the Dirac
$\gamma$-matrices.

    As this takes place, for  the calculation of amplitudes it  is
enough to make the replacement in (\ref{e3.1})
\begin {equation}
\displaystyle
\bar{Q}  \longrightarrow
{ Z \over \sqrt{ Tr ( \bar{Z} {\cal P}_i Z {\cal P}_f ) } }
\label{e3.2}
\end {equation}

    This observation allows us to supplement the computer  algebra
systems, using standard methods for calculation of cross  sections
(e.g.     \cite{st1},  \cite{st2}),   by  the   programs  for  the
calculation of the amplitudes.

    For
$
\displaystyle
Z = 1
$
    the replacement (\ref{e3.2}) takes the form
$$
\displaystyle
\bar{Q}  \longrightarrow
{ 1 \over \sqrt{ \left[ m_i m_f + ( p_i p_f ) \right]
                 \left[ 1 - ( n_i n_f ) \right]
         + ( p_i n_f ) ( p_f n_i ) } }
$$
\item[2.]
    The  expression  for  the  amplitude  of  the process contains
lesser number of the  Dirac $\gamma$-matrices than the  expression
for  the  square  of  the  modulus  of  the  amplitude of the same
process.  Because of this,  after the evaluation of traces  of the
$\gamma$-matrices the expression for the amplitude contains lesser
number of terms than the expression for the square of the  modulus
of the same amplitude.  The estimate of a gain for
$
\displaystyle
 Z = 1
$
    is  presented  in  Table  \ref{Tab1}.    (Here we consider the
processes involving massive particles.  The maximum number of  the
$\gamma$-matrices in projection operators of the massive particles
is contained in the terms
$
\displaystyle
\hat{p}_i \gamma_5 \hat{n}_i
$
    and
$
\displaystyle
\hat{p}_f \gamma_5 \hat{n}_f
\; .
$
    Therefore  the  maximum   contribution  from  the   projection
operators   in   expressions    considered    is   equal    to   4
$\gamma$-matrices.)
\begin{table}[th]
\begin{tabular}{|p{11cm}||c|c|c|}
\hline
\hline
    Number of $\gamma$-matrices in $Q \;\; ,$  $\;\; N$
& 1 & 3 & 5 \\
\hline
\hline
    Number of $\gamma$-matrices in the term of
$
\displaystyle
|M_{if}|^2
\; ,
$
    which gives the maximum contribution,
$
\displaystyle
\; \;
N_1 = 4 + 2N
$
& 6 & 10 & 14 \\
\hline
    Number of the terms after the trace evaluation,
$
\displaystyle
\; \;
(N_1 - 1)!!
$
& 15 & 945 & 135135 \\
\hline
\hline
    Number of $\gamma$-matrices in the term of
$
\displaystyle
{\cal M}_{if}
\; ,
$
    which gives the maximum contribution%
\footnotemark
,
$
\displaystyle
\; \;
N_2 = (4 + N) - 1
$
& 4 & 6 & 8 \\
\hline
    Number of the terms after the trace evaluation,
$
\displaystyle
\; \;
(N_2 - 1)!!
$
& $3$ & $15$ & $105$ \\
\hline
\hline
    The estimate of a gain,
$
\displaystyle
\; \;
{ (N_1 - 1)!!  \over (N_2 - 1)!! }
$
& 5 & 63 & 1287  \\
\hline
\hline
\end{tabular}
\caption{The estimate of a gain obtained at calculation of the
amplitudes for
$
\displaystyle
Z = 1
$
}
\label{Tab1}
\end{table}
%
\footnotetext{In this case
$
\displaystyle
(4 + N)
$
    is an odd number, and consequently the term with
$
\displaystyle
(4 + N)
\;
$
$\gamma$-matrices vanishes}

    It is follows from Table \ref{Tab1} that the more  complicated
the  process   considered  (i.e.   the  greater   number  of   the
$\gamma$-matrices is contained in the operator $Q$ which describes
the interaction) the  greater gain is  given by the  method of the
calculation  of  the  amplitude  in  comparison  with the standard
method of  the calculation  of the  square of  the modulus  of the
amplitude.  This is also true for any operator $Z$ in the  general
scheme (\ref{e2.2}).

    Really, when the number  of $\gamma$-matrices in operator  $Q$
increases by $I$, their number in the numerator of the  expression
(\ref{e2.2}) for the amplitude increases only by $I$, (denominator
does  not  change),  but  in  the  expression (\ref{e3.1}) for the
squared  amplitude  the  number  of $\gamma$-matrices increases by
$2I$ (as the formula  (\ref{e3.1}) contains both the  operator $Q$
and the operator $ \bar{Q} $).
\end{description}

    However, the scheme considered has two essential disadvantages
in the general case:
\begin{description}
\item[1.]
    There is a denominator in (\ref{e2.2}), hence the ambiguity of
the type
$
\displaystyle
{0 \over 0}
$
    can appear during the calculations. For example if one chooses
$
\displaystyle
Z = 1
$
    then the expressions for the amplitudes have the denominator
$$
\displaystyle
\sqrt{ Tr [ {\cal P}(p_i,n_i) {\cal P}(p_f,n_f) ] }
= \sqrt{ \left[ m_i m_f + ( p_i p_f ) \right]
         \left[ 1 - ( n_i n_f ) \right]
         + ( p_i n_f ) ( p_f n_i )  }
$$
    which is equal to zero for
$$
\displaystyle
n_f = - n_i + { ( p_f n_i ) \over
                    m_i m_f + ( p_i p_f ) }
  ( p_i + { m_i \over m_f } p_f )
$$
%
\item[2.]
    All the  expressions for  the amplitudes  [as it  follows from
(\ref{e2.2})] are known up to a phase factor:
\begin {equation}
\displaystyle
{\cal M}_{if} = M_{if} \cdot { \bar{u}_i Z u_f \over
                             | \bar{u}_i Z u_f | }
\;\; .
\label{e3.3}
\end {equation}

    Obviously this  fact creates  no problem, if we calculate  the
amplitude for alone diagram.  Really
\begin {equation}
\begin {array}{l} \displaystyle
( {\cal M}_{if} )^{*}
 = { [ Tr ( Q {\cal P}_i Z {\cal P}_f ) ]^{*} \over
\sqrt{ Tr ( \bar{Z} {\cal P}_i Z {\cal P}_f ) } }
= { [ Tr ( Q u_i \bar{u}_i Z u_f \bar{u}_f ) ]^{*} \over
\sqrt{ Tr ( \bar{Z} {\cal P}_i Z {\cal P}_f ) } }
 = { [ ( \bar{u}_f Q u_i ) ( \bar{u}_i Z u_f ) ]^{*} \over
\sqrt{ Tr ( \bar{Z} {\cal P}_i Z {\cal P}_f ) } }
                        \\[0.5cm] \displaystyle
 = { ( \bar{u}_f \bar{Z} u_i ) ( \bar{u}_i \bar{Q} u_f ) \over
\sqrt{ Tr ( \bar{Z} {\cal P}_i Z {\cal P}_f ) } }
   = { Tr ( \bar{Z} u_i \bar{u}_i \bar{Q} u_f \bar{u}_f ) \over
\sqrt{ Tr ( \bar{Z} {\cal P}_i Z {\cal P}_f ) } }
   = { Tr ( \bar{Z} {\cal P}_i \bar{Q} {\cal P}_f ) \over
\sqrt{ Tr ( \bar{Z} {\cal P}_i Z {\cal P}_f ) } }
\;\; ,
\end {array}
\label{e3.4}
\end {equation}
\begin {equation}
\begin {array}{c} \displaystyle
{\cal M}_{if} ( {\cal M}_{if} )^{*}
= { Tr ( Q {\cal P}_i Z {\cal P}_f ) \cdot
    Tr ( \bar{Z} {\cal P}_i \bar{Q} {\cal P}_f ) \over
    Tr ( \bar{Z} {\cal P}_i Z {\cal P}_f ) }
                        \\[0.5cm] \displaystyle
= { Tr ( Q {\cal P}_i \bar{Q} {\cal P}_f ) \cdot
    Tr ( \bar{Z} {\cal P}_i Z {\cal P}_f ) \over
    Tr ( \bar{Z} {\cal P}_i Z {\cal P}_f ) }
  = Tr ( Q {\cal P}_i \bar{Q} {\cal P}_f ) = | M_{if} |^2
\;\; .
\end {array}
\label{e3.5}
\end {equation}
    [In (\ref{e3.5}) we used the identity (\ref{e2.7}).]

    However,  if  we  have  the  interfering  diagrams,  then  the
application of (\ref{e2.2}) leads to the fact that the expressions
for  the  amplitudes  corresponding  to  different  channels   are
multiplied by  different phase  factors in  the general  case.  In
this situation the additional  calculations are necessary.   These
calculations are considered below.
\end{description}

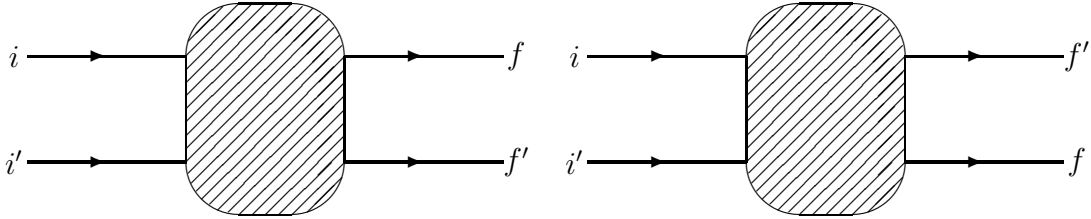
\begin{figure}[ht]
\begin{tabular}{cc}
\begin{picture}(200,100)
\put(100,50){\oval(60,80)}
\put(71,76){\line(1,1){13.}}
\put(70,70){\line(1,1){20.}}
\put(70,65){\line(1,1){25.}}
\put(70,60){\line(1,1){30.}}
\put(70,55){\line(1,1){35.}}
\put(70,50){\line(1,1){40.}}
\put(70,45){\line(1,1){44.}}
\put(70,40){\line(1,1){48.}}
\put(70,35){\line(1,1){51.}}
\put(70,30){\line(1,1){54.}}
\put(71,26){\line(1,1){55.}}
\put(72,22){\line(1,1){56.}}
\put(74,19){\line(1,1){55.}}
\put(76,16){\line(1,1){54.3}}
\put(79,14){\line(1,1){51.}}
\put(82,12){\line(1,1){48.}}
\put(86,11){\line(1,1){44.}}
\put(90,10){\line(1,1){40.}}
\put(95,10){\line(1,1){35.}}
\put(100,10){\line(1,1){30.}}
\put(105,10){\line(1,1){25.}}
\put(110,10){\line(1,1){20.}}
\put(116,11){\line(1,1){13.}}
\put(05,30){\makebox(0,0){$ i' $}}
\put(05,70){\makebox(0,0){$ i $}}
\put(195,30){\makebox(0,0){$ f' $}}
\put(195,70){\makebox(0,0){$ f $}}
\thicklines
\put(10,30){\line(2,0){60.}}
\thicklines
\put(10,70){\line(2,0){60.}}
\thicklines
\put(130,30){\line(2,0){60.}}
\thicklines
\put(130,70){\line(2,0){60.}}
\put(40,30){\vector(1,0){0.}}
\put(40,70){\vector(1,0){0.}}
\put(160,30){\vector(1,0){0.}}
\put(160,70){\vector(1,0){0.}}
\end{picture}
&
\begin{picture}(200,100)
\put(100,50){\oval(60,80)}
\put(71,76){\line(1,1){13.}}
\put(70,70){\line(1,1){20.}}
\put(70,65){\line(1,1){25.}}
\put(70,60){\line(1,1){30.}}
\put(70,55){\line(1,1){35.}}
\put(70,50){\line(1,1){40.}}
\put(70,45){\line(1,1){44.}}
\put(70,40){\line(1,1){48.}}
\put(70,35){\line(1,1){51.}}
\put(70,30){\line(1,1){54.}}
\put(71,26){\line(1,1){55.}}
\put(72,22){\line(1,1){56.}}
\put(74,19){\line(1,1){55.}}
\put(76,16){\line(1,1){54.3}}
\put(79,14){\line(1,1){51.}}
\put(82,12){\line(1,1){48.}}
\put(86,11){\line(1,1){44.}}
\put(90,10){\line(1,1){40.}}
\put(95,10){\line(1,1){35.}}
\put(100,10){\line(1,1){30.}}
\put(105,10){\line(1,1){25.}}
\put(110,10){\line(1,1){20.}}
\put(116,11){\line(1,1){13.}}
\put(05,30){\makebox(0,0){$ i' $}}
\put(05,70){\makebox(0,0){$ i $}}
\put(195,30){\makebox(0,0){$ f $}}
\put(195,70){\makebox(0,0){$ f' $}}
\thicklines
\put(10,30){\line(2,0){60.}}
\thicklines
\put(10,70){\line(2,0){60.}}
\thicklines
\put(130,30){\line(2,0){60.}}
\thicklines
\put(130,70){\line(2,0){60.}}
\put(40,30){\vector(1,0){0.}}
\put(40,70){\vector(1,0){0.}}
\put(160,30){\vector(1,0){0.}}
\put(160,70){\vector(1,0){0.}}
\end{picture}
\end{tabular}
\caption{The general form of the diagrams of the process
proceeding in two interfering channels}
\label{Fg1}
\end{figure}

    Let us  consider the  diagrams of  the process  of the general
form,  which  proceeds  in   two  different  channels  (see   Fig.
\ref{Fg1}).

    The expression
\begin {equation}
\displaystyle
M = ( \bar{u}_f Q u_i ) \cdot ( \bar{u}_{f'} R u_{i'} ) = M_{if}
\cdot M_{i' f'}
\;\;  ,
\label{e3.6}
\end {equation}
    corresponds to the first diagram. The expression
\begin {equation}
\displaystyle
 M'= ( \bar{u}_{f'} S u_i ) \cdot ( \bar{u}_f T u_{i'} )
            = M_{i f'} \cdot M_{i' f}
\label{e3.7}
\end {equation}
    corresponds to the  second one, where  $Q$, $R$, $S$,  $T$ are
arbitrary matrix operators characterizing the interaction.

    There are  three possibilities  to calculate  the amplitude of
this process correctly:
%
\begin{description}
\item[(a)]
    One can apply the Fierz transformation for the second diagram:
$
\displaystyle
f \leftrightarrow f'
$.
    However  this  method  requires  a  large number of additional
calculations.

\item[(b)]
    One can  multiply both  of the  amplitudes by  the same  phase
factors, for example by
$$
\displaystyle
{ \bar{u}_i Z_1 u_f \over | \bar{u}_i Z_1 u_f | } \cdot
{ \bar{u}_{i'} Z_2 u_{f'} \over | \bar{u}_{i'} Z_2 u_{f'} | }
\;\; .
$$
    In this case we have the expression for the first diagram
$$
\displaystyle
     { Tr ( Q {\cal P}_i Z_1 {\cal P}_f ) \over
\sqrt{ Tr ( \bar{Z}_1 {\cal P}_i Z_1 {\cal P}_f ) } }
\cdot
     { Tr ( R {\cal P}_{i'} Z_2 {\cal P}_{f'} ) \over
\sqrt{ Tr ( \bar{Z}_2  {\cal P}_{i'} Z_2 {\cal P}_{f'} ) }
}
$$
    and for the second diagram
$$
\displaystyle
{ Tr ( S {\cal P}_i Z_1 {\cal P}_f T {\cal P}_{i'}
     Z_2 {\cal P}_{f'} ) \over
\sqrt{ Tr ( \bar{Z}_1 {\cal P}_i Z_1 {\cal P}_f ) }
 \cdot
\sqrt{ Tr ( \bar{Z}_2 {\cal P}_{i'} Z_2 {\cal P}_{f'} ) }
}
\;\; .
$$
    It is obvious  that the calculation  of the amplitude  for the
second  diagram  is  complicated  enough.    Besides,  it  is  not
convenient to use the  expressions which have different  structure
for the calculation of amplitudes for the different diagrams.
\item[(c)]
    The third possibility is  to calculate the relative  phase for
the first and second diagrams  and to use the expression  obtained
for the phase correction of one of two amplitudes.
\end{description}
    We will use the third possibility. Interference term has the
form
$$
\displaystyle
M \cdot ( M')^{*}
 = M_{if} \cdot M_{i' f'} \cdot ( M_{i f'} )^{*}
 \cdot ( M_{i' f} )^{*}
 = ( \bar{u}_f Q u_i ) ( \bar{u}_{f'} R u_{i'} )
   ( \bar{u}_i \bar{S} u_{f'} )
   ( \bar{u}_{i'} \bar{T} u_f )
\;\; .
$$
    Let us multiply the interference term by
$$
\displaystyle
   { ( \bar{u}_i Z_1 u_f )
     ( \bar{u}_{i'} Z_2 u_{f'} )
     ( \bar{u}_{f'} \bar{Z}_3 u_i )
     ( \bar{u}_f \bar{Z}_4 u_{i'} ) \over
     ( \bar{u}_i Z_1 u_f )
     ( \bar{u}_{i'} Z_2 u_{f'} )
     ( \bar{u}_{f'} \bar{Z}_3 u_i )
     ( \bar{u}_f \bar{Z}_4 u_{i'} ) }
\cdot { ( \bar{u}_f \bar{Z}_1 u_i )
        ( \bar{u}_i Z_3 u_{f'} )
        ( \bar{u}_{f'} \bar{Z}_2 u_{i'} )
        ( \bar{u}_{i'} Z_4 u_f ) \over
        ( \bar{u}_f \bar{Z}_1 u_i )
        ( \bar{u}_i Z_3 u_{f'} )
        ( \bar{u}_{f'} \bar{Z}_2 u_{i'} )
        ( \bar{u}_{i'} Z_4 u_f ) }  \equiv 1
$$
    where
$ \; Z_1$, $Z_2$, $Z_3$, $Z_4 \; $
    are still arbitrary matrix operators.

    As a result in this case we have
\begin {equation}
\begin {array}{c} \displaystyle
M \cdot ( M')^{*} \equiv
  { Tr ( Q {\cal P}_i Z_1 {\cal P}_f ) \over
    Tr ( \bar{Z}_1 {\cal P}_i Z_1 {\cal P}_f ) } \cdot
  { Tr ( R {\cal P}_{i'} Z_2 {\cal P}_{f'}) \over
    Tr ( \bar{Z}_2 {\cal P}_{i'} Z_2 {\cal P}_{f'} ) }
             \\           \\     \displaystyle
\times { Tr ( \bar{Z}_3 {\cal P}_i \bar{S} {\cal P}_{f'} )
                 \over
    Tr ( \bar{Z}_3 {\cal P}_i Z_3 {\cal P}_{f'} ) } \cdot
  { Tr ( \bar{Z}_4 {\cal P}_{i'} \bar{T} {\cal P}_f )
                 \over
      Tr ( \bar{Z}_4 {\cal P}_{i'} Z_4 {\cal P}_f ) }
\cdot Tr ( \bar{Z}_1 {\cal P}_i Z_3 {\cal P}_{f'}
           \bar{Z}_2 {\cal P}_{i'} Z_4 {\cal P}_f )
              \\          \\    \displaystyle
 = { Tr ( Q {\cal P}_i Z_1 {\cal P}_f ) \over
\sqrt{ Tr ( \bar{Z}_1 {\cal P}_i Z_1 {\cal P}_f ) } }
  \cdot
  { Tr ( R {\cal P}_{i'} Z_2 {\cal P}_{f'} ) \over
\sqrt
{ Tr (\bar{Z}_2 {\cal P}_{i'} Z_2 {\cal P}_{f'} ) } }
\cdot { Tr ( \bar{Z}_3 {\cal P}_i \bar{S} {\cal P}_{f'} )
                  \over
 \sqrt{ Tr ( \bar{Z}_3 {\cal P}_i Z_3 {\cal P}_{f'} ) } }
\cdot { Tr ( \bar{Z}_4 {\cal P}_{i'} \bar{T} {\cal P}_f )
\over
\sqrt{ Tr ( \bar{Z}_4 {\cal P}_{i'} Z_4 {\cal P}_f ) } }
              \\          \\        \displaystyle
\times { Tr ( \bar{Z}_1 {\cal P}_i Z_3 {\cal P}_{f'}
              \bar{Z}_2 {\cal P}_{i'} Z_4 {\cal P}_f )
\over
\sqrt{ Tr ( \bar{Z}_1 {\cal P}_i Z_1 {\cal P}_f )
       Tr ( \bar{Z}_3 {\cal P}_i Z_3 {\cal P}_{f'} )
       Tr ( \bar{Z}_2 {\cal P}_{i'} Z_2 {\cal P}_{f'} )
       Tr ( \bar{Z}_4 {\cal P}_{i'} Z_4 {\cal P}_f ) } }
              \\          \\        \displaystyle
= {\cal M}_{if} \cdot
{\cal M}_{i' f'} \cdot ( {\cal M}_{i f'} )^{*}
    \cdot ( {\cal M}_{i' f} )^{*} \cdot K
\end {array}
\label{e3.8}
\end {equation}
    where
$ \; {\cal M}_{if}$, ${\cal M}_{i' f'}$,
$( {\cal M}_{i f'} )^{*}$, $( {\cal M}_{i' f} )^{*}$
    are  given  by  expressions  analogous  to  (\ref{e2.2})   and
(\ref{e3.4}), and the coefficient $K$ is given by
\begin {equation}
\displaystyle
K= { Tr ( \bar{Z}_1 {\cal P}_i Z_3 {\cal P}_{f'}
          \bar{Z}_2 {\cal P}_{i'} Z_4 {\cal P}_f ) \over
\sqrt{ Tr ( \bar{Z}_1 {\cal P}_i Z_1 {\cal P}_f )
       Tr ( \bar{Z}_3 {\cal P}_i Z_3 {\cal P}_{f'} )
       Tr ( \bar{Z}_2 {\cal P}_{i'} Z_2 {\cal P}_{f'} )
       Tr ( \bar{Z}_4 {\cal P}_{i'} Z_4 {\cal P}_f ) } }
.
\label{e3.9}
\end {equation}
    It is obvious that
$$
\displaystyle
|K| = 1  \, .
$$
    Thus we have to calculate the amplitude of process with
interfering diagrams as follows
\begin {equation}
\displaystyle
{\cal M} + {\cal M}' =
 K \cdot {\cal M}_{if} \cdot {\cal M}_{i' f'} +
         {\cal M}_{i f'} \cdot {\cal M}_{i' f}
\;\; .
\label{e3.10}
\end {equation}
    In particularly, if
$$
\displaystyle
Z_1 = Z_2 = Z_3 = Z_4 = 1
$$
    we have
$$
\displaystyle
K= { Tr ( {\cal P}_i {\cal P}_{f'}
          {\cal P}_{i'} {\cal P}_f ) \over
\sqrt{ Tr ( {\cal P}_i {\cal P}_f )
       Tr ( {\cal P}_i {\cal P}_{f'} )
       Tr ( {\cal P}_{i'} {\cal P}_{f'} )
       Tr ( {\cal P}_{i'} {\cal P}_f ) } }
$$
    In the massless limit we have
\begin {equation}
\displaystyle
\bar{u}_{\pm}(p_f) Q u_{\mp}(p_i) \simeq
{ Tr [ Q \hat{p}_i \hat{p}_f ( 1 \mp \gamma _5 ) ]
\over 2 \sqrt{ 2 ( p_i p_f ) } }
\; ,
\label{e3.11}
\end {equation}
\begin {equation}
\displaystyle
 K = { Tr [ \hat{p}_i \hat{p}_{f'} \hat{p}_{i'}
\hat{p}_f ( 1 \mp \gamma_5 ) ] \over 8
\sqrt{ ( p_i p_f ) ( p_i p_{f'} ) ( p_{i'} p_f ) ( p_{i'} p_{f'} ) } }
\; .
\label{e3.12}
\end {equation}
    The  formulae  (\ref{e3.11}),  (\ref{e3.12})  generalize   the
method of the calculation of amplitudes proposed in \cite{r2.1-6}.


\section {A new method of the calculation of amplitudes}

    All the disadvantages of the general scheme of the calculation
of the amplitudes listed in the previous section can be eliminated
with the help of a successful choice of the operator $Z$.

    As a suitable choice we propose \\
\hspace*{30mm}
$ Z = {\cal P} \;\;\; $ [see (\ref{e2.3})] $\;\;$
\hspace{2mm} or \hspace{5mm}
$Z = {\cal P}_{\pm} \;\;\; $ [see (\ref{e2.4})] .  \\
    Really, in this case the unknown phase factor for each line of
the diagram splits into two parts [see (\ref{e3.3})]:
$$
\displaystyle
{\cal M}_{if}
= M_{if} \cdot { \bar{u}_i {\cal P} u_f \over
               | \bar{u}_i {\cal P} u_f | }
= M_{if} \cdot { \bar{u}_i u \bar{u} u_f \over
               | \bar{u}_i u \bar{u} u_f | }
= M_{if} \cdot { \bar{u}_i u \over | \bar{u}_i u | }
         \cdot { \bar{u} u_f \over | \bar{u} u_f | }
\;\; .
$$
   Let us consider again the diagrams of the process shown
in Fig. \ref{Fg1}.

    Calculating the amplitudes (\ref{e3.6}), (\ref{e3.7}),
we have
$$
\begin {array}{c} \displaystyle
{\cal M} = M_{if} \cdot M_{i' f'}
          \cdot { \bar{u}_i u  \over  | \bar{u}_i u | }
          \cdot { \bar{u} u_f  \over  | \bar{u} u_f | }
          \cdot { \bar{u}_{i'} u \over  | \bar{u}_{i'} u | }
          \cdot { \bar{u} u_{f'} \over  | \bar{u} u_{f'} | }
\;\; ,
                        \\[0.5cm] \displaystyle
{\cal M}' = M_{i f'} \cdot M_{i' f}
        \cdot { \bar{u}_i u  \over  | \bar{u}_i u | }
        \cdot { \bar{u} u_{f'} \over  | \bar{u} u_{f'} | }
        \cdot { \bar{u}_{i'} u \over  | \bar{u}_{i'} u | }
        \cdot { \bar{u} u_f  \over  | \bar{u} u_f | }
\;\; ,
\end {array}
$$
    that  is  the  amplitudes   of  the  different  diagrams   are
multiplied by the same phase factor, and therefore one can  ignore
it.  This  conclusion is right  for any number  of the interfering
diagrams and for any number of the nonclosed fermion lines in  the
diagrams.

    We also note that, as would be expected [see~(\ref{e3.9})]
$$
 K \equiv 1
$$
    if
$$
Z_1 = Z_2 = Z_3 = Z_4 = {\cal P}
$$
    or

$$
Z_1 = Z_2 = Z_3 = Z_4 = {\cal P}_{\pm}
$$
    since projection operators have the following properties:
\begin {equation}
\displaystyle
\bar{\cal P}={\cal P}
\;\; ,   \;\;\;\;\;
{\cal P} A {\cal P} = Tr [ {\cal P} A ] \cdot {\cal P}
\;\; ,
\label{e4.1}
\end {equation}
\begin {equation}
\displaystyle
\bar{\cal P}_{\pm} = {\cal P}_{\pm}
\;\; , \;\;\;\;\;
{\cal P}_{\pm} A {\cal P}_{\pm}
= Tr [ {\cal P}_{\pm} A ] \cdot {\cal P}_{\pm}
\;\;\;  .
\label{e4.2}
\end {equation}
    Really
$$
\begin {array}{l}
\displaystyle
\bar{\cal P} = \gamma^0 {\cal P}^{+} \gamma^0
= \gamma^0 ( u \bar{u} )^{+} \gamma^0
= \gamma^0 ( u u^{+} \gamma^0 )^{+} \gamma^0
                        \\[0.5cm] \displaystyle
= \gamma^0 [ ( \gamma^0 )^{+} ( u^{+} )^{+} u^{+} ] \gamma^0
= \gamma^0 [ \gamma^0 u u^{+} ] \gamma^0
= u u^{+} \gamma^0 = u \bar{u} = {\cal P}
\;\;  ,
\end {array}
$$
$$
\begin  {array}{l}
\displaystyle
{\cal P} A {\cal P} = (u)_{\alpha} (\bar{u})_{\beta}
  (A)^{\beta \rho} (u)_{\rho} (\bar{u})_{\delta}
= [ (\bar{u})_{\beta} (A)^{\beta \rho} (u)_{\rho} ]
  (u)_{\alpha} (\bar{u})_{\delta}
                        \\[0.5cm] \displaystyle
= [ (u)_{\rho } (\bar{u})_{\beta} (A)^{\beta \rho} ]
  (u)_{\alpha} (\bar{u})_{\delta}
= Tr [ {\cal P}  A ] \cdot {\cal P}
\; .
\end {array}
$$

    Therefore we can calculate the amplitude of the process with
interfering diagrams as follows
\begin {equation}
\begin {array}{r} \displaystyle
M + M'=
 { Tr ( Q {\cal P}_i {\cal P} {\cal P}_f ) \over
\sqrt{ Tr ( {\cal P} {\cal P}_i ) Tr ( {\cal P} {\cal P}_f ) } }
                       \cdot
 { Tr ( R {\cal P}_{i'} {\cal P} {\cal P}_{f'} ) \over
\sqrt{ Tr ( {\cal P} {\cal P}_{i'} ) Tr ( {\cal P} {\cal P}_{f'} ) } }
   \\         \\        \displaystyle
+ { Tr ( S {\cal P}_i {\cal P} {\cal P}_{f'} ) \over
\sqrt{ Tr ( {\cal P} {\cal P}_i ) Tr ( {\cal P} {\cal P}_{f'} ) } }
                       \cdot
 { Tr ( T {\cal P}_{i'} {\cal P} {\cal P}_f ) \over
\sqrt{ Tr ( {\cal P} {\cal P}_{i'} ) Tr ( {\cal P} {\cal P}_f ) } }
   \\         \\        \displaystyle
= { Tr ( Q {\cal P}_i {\cal P} {\cal P}_f ) \cdot
    Tr ( R {\cal P}_{i'} {\cal P} {\cal P}_{f'} )
  + Tr ( S {\cal P}_i {\cal P} {\cal P}_{f'} ) \cdot
    Tr ( T {\cal P}_{i'} {\cal P} {\cal P}_f ) \over
\sqrt{ Tr ( {\cal P} {\cal P}_i ) Tr ( {\cal P} {\cal P}_{i'} )
       Tr ( {\cal P} {\cal P}_f ) Tr ( {\cal P} {\cal P}_{f'} ) } }
\; .
\label{e4.3}
\end {array}
\end {equation}
    [where  for  the  denominators  in  (\ref{e4.3})  we  used the
identity
$$
\displaystyle
  Tr ( \bar{\cal P} {\cal P}_i {\cal P} {\cal P}_f )
= Tr ( {\cal P} {\cal P}_i) Tr({\cal P} {\cal P}_f )
\; ,
$$
    which follows from (\ref{e4.1})~--~(\ref{e4.2})].

    This  expression  enables   us  to  calculate   the  amplitude
numerically.  The complex numbers obtained in this way may be used
to calculate the squared amplitude.

    Note that in (\ref{e4.3})  and in the analogous  formulae used
further we  shall use  the  equality  sign instead  of the  symbol
$ \simeq $, since now there does  not exist  any trouble  with the
phase factors.  (However one can not do this when the phase factor
is not the same for all terms!  -- see, for example, subsection  5
of Section~{\bf 6}.)

    Thus,  if  we  use  the  projection  operator  for $Z$ then in
addition to the two advantages of the general  scheme (\ref{e2.2})
considered in Section {\bf3} we obtain one more advantage:
%
\begin{description}
\item[3.]
    For the processes with interfering diagrams the number of  the
expressions which are necessary to be calculated is reduced  since
it is not necessary to calculate the interfering terms.
\end{description}

    If  in  individual  cases  under  numerical  calculations  the
denominator in (\ref{e4.3}) vanishes,  it is sufficient to  change
an arbitrary projection operator ${\cal P}$ entering  (\ref{e4.3})
(i.e. to change  arbitrary 4-vectors entering  it).  In  this case
the amplitudes for  the different diagrams  are multiplied by  one
and the same phase factor.  Really, under the replacement of
$
\;
\displaystyle
{\cal P}
\;
$
by
$
\;
\displaystyle
{\cal P}'
\; ,
$
    we obtain [using  (\ref{e2.5}) and (\ref{e2.6})]:
$$
\displaystyle
{\cal M}_{if} = { Tr ( Q {\cal P}_i {\cal P} {\cal P}_f )
\over
\sqrt{ Tr ( {\cal P} {\cal P}_i ) Tr ( {\cal P} {\cal P}_f ) } }
   = { Tr ( Q {\cal P}_i {\cal P}' {\cal P}_f ) \over
\sqrt{ Tr ( {\cal P}' {\cal P}_i ) Tr ( {\cal P}' {\cal P}_f ) } }
\cdot
\left[ { Tr ( {\cal P}' {\cal P}_i {\cal P} {\cal P}_f ) \over
         Tr ( {\cal P} {\cal P}_i {\cal P}' {\cal P}_f ) }
                                                \right] ^{1/2} ,
$$
$$
\displaystyle
{\cal M}_{i' f'} = { Tr ( R {\cal P}_{i'}
                        {\cal P} {\cal P}_{f'} ) \over
\sqrt{ Tr ( {\cal P} {\cal P}_{i'} )
       Tr ( {\cal P} {\cal P}_{f'} ) } }
   = { Tr ( R {\cal P}_{i'} {\cal P}'
              {\cal P}_{f'} ) \over
\sqrt{ Tr ( {\cal P}' {\cal P}_{i'} )
       Tr ( {\cal P}' {\cal P}_{f'} ) } } \cdot
\left[ { Tr ( {\cal P}' {\cal P}_{i'} {\cal P}
              {\cal P}_{f'} ) \over
  Tr ( {\cal P} {\cal P}_{i'} {\cal P}'
       {\cal P}_{f'} ) } \right] ^{1/2} ,
$$
$$
\displaystyle
{\cal M}_{i f'} = { Tr ( S {\cal P}_i {\cal P}
                                 {\cal P}_{f'} ) \over
\sqrt{ Tr ( {\cal P} {\cal P}_i )
       Tr ( {\cal P} {\cal P}_{f'} ) } }
   = { Tr ( S {\cal P}_i {\cal P}' {\cal P}_{f'} ) \over
\sqrt{ Tr ( {\cal P}' {\cal P}_i )
       Tr ( {\cal P}' {\cal P}_{f'} ) } } \cdot
\left[ { Tr ( {\cal P}' {\cal P}_i
              {\cal P} {\cal P}_{f'} ) \over
         Tr ( {\cal P} {\cal P}_i {\cal P}'
              {\cal P}_{f'} ) } \right] ^{1/2} ,
$$
$$
\displaystyle
{\cal M}_{i' f} = { Tr ( T {\cal P}_{i'}
                                 {\cal P} {\cal P}_f ) \over
\sqrt{ Tr ( {\cal P} {\cal P}_{i'} )
       Tr ( {\cal P} {\cal P}_f ) } }
   = { Tr ( T {\cal P}_{i'} {\cal P}' {\cal P}_f ) \over
\sqrt{ Tr ( {\cal P}' {\cal P}_{i'} )
       Tr ( {\cal P}' {\cal P}_f ) } } \cdot
\left[ { Tr ( {\cal P}' {\cal P}_{i'} {\cal P} {\cal P}_f ) \over
  Tr ( {\cal P} {\cal P}_{i'} {\cal P}'
       {\cal P}_f ) } \right] ^{1/2} .
$$
    As this takes place we have [see~(\ref{e2.7})]
$$
\displaystyle
\left[ { Tr ( {\cal P}' {\cal P}_i {\cal P} {\cal P}_f ) \over
  Tr ( {\cal P} {\cal P}_i {\cal P}' {\cal P}_f ) }
\cdot
 { Tr ( {\cal P}' {\cal P}_{i'} {\cal P}
        {\cal P}_{f'} ) \over
   Tr ( {\cal P} {\cal P}_{i'} {\cal P}'
        {\cal P}_{f'} ) } \right] ^{1/2} \equiv
\left[ { Tr ( {\cal P}'{\cal P}_i {\cal P}
              {\cal P}_{f'} ) \over
         Tr ( {\cal P} {\cal P}_i {\cal P}'
              {\cal P}_{f'} ) } \cdot
 { Tr ( {\cal P}' {\cal P}_{i'} {\cal P} {\cal P}_f )
\over
  Tr ( {\cal P} {\cal P}_{i'} {\cal P}' {\cal P}_f ) }
\right] ^{1/2} .
$$
    Note that  if the  operators, characterizing  the interaction,
contain the product  of $\gamma$-matrices the  number of which  is
greater  than  the  number  of  $\gamma$-matrices  involved in the
projection operator, then the calculation of  the  amplitude for a
single  diagram  is  easer  than  the  calculation  of the squared
amplitude.

    However,  for  the  processes  with  interfering  diagrams the
method of  the calculation  of amplitudes  is easier  in any case,
because we need not calculate the interference terms.

    As  it  was  already  mentioned,  the method considered can be
generalized easily to the case of the reaction with  participation
of antiparticles.  To do  this it is sufficient to  substitute the
projection operators of antiparticles in place of the operators of
particles in (\ref{e2.2}).  For  example, if we are interested  in
the value
$
\displaystyle
\bar{v}_f Q u_i
$ ,
    where $v_f$ is the bispinor of a free antiparticle, then
\begin {equation}
\displaystyle
\bar{v}_f Q u_i = { Tr ( Q u_i \bar{u}_i Z v_f \bar{v}_f )
\over
\sqrt { Tr ( \bar{Z} u_i \bar{u}_i Z v_f \bar{v}_f ) } }
\label{e4.4}
\end {equation}
    where
\begin {equation}
\displaystyle
v(p,n) \bar{v}(p,n)
= { 1 \over 2 }( -m + \hat{p} ) ( 1+\gamma_5 \hat{n} )
\label{e4.5}
\end {equation}
$$
\displaystyle
p^2 = m^2
\; , \;\;\;
n^2 = -1
\; , \;\;\;
pn = 0
\; , \;\;
\bar{v} v = - 2 m
$$
    for a massive antiparticle, or
\begin {equation}
\displaystyle
v_{\pm}(p) \bar{v}_{\pm}(p)
= { 1 \over 2 } ( 1 \mp \gamma_5 ) \hat{p}
\label{e4.6}
\end {equation}
$$
\displaystyle
p^2 = 0
\; , \;\;
\bar{v}_{\pm} \gamma_{\mu} v_{\pm} =  2 p_{\mu}
$$
    for a massless antiparticle.
\newline
    As always, we use (\ref{e2.3}) or (\ref{e2.4}) instead of $Z$.


    \section  {The formulae for the calculation of the amplitudes
of the processes involving the polarized Dirac particles}

    In  this  Section  the  expressions  for the amplitudes of the
processes with both the  massive and massless Dirac  particles are
presented.   Using the  formulae (\ref{e2.2})~--~(\ref{e2.4})  and
(\ref{e4.4})~--~(\ref{e4.6}), we obtain:
\begin {equation}
\begin {array}{l} \displaystyle
  \bar{u}_{\pm}(p_f) Q u_{\pm}(p_i)
= \bar{u}_{\pm}(p_f) Q v_{\mp }(p_i)
= \bar{v}_{\mp}(p_f) Q u_{\pm }(p_i)
= \bar{v}_{\mp}(p_f) Q v_{\mp }(p_i)
                        \\[0.5cm] \displaystyle
= { Tr [ Q \hat{p}_i \hat{q} \hat{p}_f ( 1 \mp \gamma_5 ) ] \over
   4 \sqrt{ ( q p_i ) ( q p_f ) } }
\;\;\; .
\end {array}
\label{e5.1}
\end {equation}
    Here
$$
\displaystyle
Z = { 1 \over 2 } ( 1 \mp \gamma_5) \hat{q} = {\cal P}_{\mp}
\; , \;\;
q^2 = 0 \, .
$$
    The massless 4-vector $q$ can be arbitrary, but it must be the
same for all nonclosed fermion lines of the diagrams considered.
\begin {equation}
\begin {array}{l} \displaystyle
  \bar{u}_{\pm}(p_f) Q u_{\mp}(p_i)
= \bar{u}_{\pm}(p_f) Q v_{\pm}(p_i)
= \bar{v}_{\mp}(p_f) Q u_{\mp}(p_i)
= \bar{v}_{\mp}(p_f) Q v_{\pm}(p_i)
                        \\[0.5cm] \displaystyle
= { Tr [ Q \hat{p}_i
( m \mp \hat{r} \hat{l} ) \hat{p}_f ( 1 \mp \gamma_5 ) ] \over
4 \sqrt{ [ ( r p_i ) \pm m ( l p_i ) ]
         [ ( r p_f ) \mp m ( l p_f ) ] } }
\;\; .
\end {array}
\label{e5.2}
\end {equation}
    Here
$$
\displaystyle
Z = { 1 \over  2 } ( m + \hat{r} ) ( 1 + \gamma_5 \hat{l} )
= {\cal P} ,
\;\;\;
r^2 = m^2 , \;\; l^2 = -1 ,  \;\; rl = 0  \, .
$$
    The same remark as for the vector $q$ in (\ref{e5.1}) is takes
place for the 4-vectors $r$ and $l$.

    In the latter case we can not use the easier operator
$
\displaystyle
{\cal P}_{\pm}
$
    for $Z$, since in this case the numerator and the  denominator
are equal to  zero.  However  to simplify the  calculations we can
require
$
\displaystyle
m = 0, \;\;\; r^2 = 0 \;\;
$
    in (\ref{e5.2}):
\begin {equation}
\begin {array}{l} \displaystyle
  \bar{u}_{\pm}(p_f) Q u_{\mp}(p_i)
= \bar{u}_{\pm}(p_f) Q v_{\pm}(p_i)
= \bar{v}_{\mp}(p_f) Q u_{\mp}(p_i)
= \bar{v}_{\mp}(p_f) Q v_{\pm}(p_i)
                        \\[0.5cm] \displaystyle
= \mp { Tr [ Q \hat{p}_i
 \hat{r} \hat{l} \hat{p}_f ( 1 \mp \gamma_5 ) ] \over
4 \sqrt{ ( r p_i ) ( r p_f ) } }
\;\;\; .
\end {array}
\label{e5.3}
\end {equation}
    This  formula  generalizes  the  method  of the calculation of
amplitudes proposed in \cite{r2.10a}~--~\cite{r2.10b}.
     Further
\begin {equation}
\displaystyle
  \bar{u}_{\pm}(p_f) Q u(p_i,n_i)
= \bar{v}_{\mp}(p_f) Q u(p_i,n_i)
= { Tr \left[ Q ( m_i \pm \hat{p}_i \hat{n}_i
  + \hat{p}_i \pm m_i \hat{n}_i )
 \hat{q} \hat{p}_f ( 1 \mp \gamma_5 ) \right] \over
4 \sqrt{ 2 \left[ ( q p_i ) \pm  m_i ( q n_i ) \right] ( q p_f ) } }
\;\;\; ,
\label{e5.4}
\end {equation}
\begin {equation}
\displaystyle
  \bar{u}_{\pm}(p_f) Q v(p_i,n_i)
= \bar{v}_{\mp}(p_f) Q v(p_i,n_i)
= { Tr \left[ Q ( - m_i \pm \hat{p}_i \hat{n}_i
  + \hat{p}_i \mp m_i \hat{n}_i )
 \hat{q} \hat{p}_f ( 1 \mp \gamma_5 ) \right] \over
4 \sqrt { 2 \left[ ( q p_i ) \mp  m_i ( q n_i ) \right] ( q p_f ) } }
\;\;\; ,
\label{e5.5}
\end {equation}
\begin {equation}
\displaystyle
  \bar{u}(p_f,n_f) Q u_{\pm}(p_i)
= \bar{u}(p_f,n_f) Q v_{\mp}(p_i)
= { Tr \left[ Q  (1 \pm \gamma_5 ) \hat{p}_i \hat{q}
( m_f \mp \hat{p}_f \hat{n}_f
  + \hat{p}_f \pm m_f \hat{n}_f ) \right] \over
4 \sqrt{ 2 ( q p_i )
\left[ ( q p_f ) \pm  m_f ( q n_f ) \right]  } }
\;\;\; ,
\label{e5.6}
\end {equation}
\begin {equation}
\displaystyle
  \bar{v}(p_f,n_f) Q u_{\pm }(p_i)
= \bar{v}(p_f,n_f) Q v_{\mp }(p_i)
= { Tr \left[ Q  ( 1 \pm \gamma_5 ) \hat{p}_i \hat{q}
( - m_f \mp \hat{p}_f \hat{n}_f
  + \hat{p}_f \mp m_f \hat{n}_f ) \right] \over
4 \sqrt{ 2 ( q p_i )
\left[ ( q p_f ) \mp  m_f ( q n_f ) \right] } }
\;\;\; ,
\label{e5.7}
\end {equation}
\begin {equation}
\begin {array}{l} \displaystyle
  \bar{u}(p_f,n_f) Q u(p_i,n_i)
                        \\[0.5cm] \displaystyle
= { Tr \left[ Q
( m_i \pm \hat{p}_i \hat{n}_i
  + \hat{p}_i \pm m_i \hat{n}_i )
  ( 1 \mp \gamma_5 ) \hat{q}
( m_f \mp \hat{p}_f \hat{n}_f
  + \hat{p}_f \pm m_f \hat{n}_f ) \right] \over
8 \sqrt{ \left[ ( q p_i ) \pm  m_i ( q n_i ) \right]
\left[ ( q p_f ) \pm  m_f ( q n_f ) \right] } }
\;\;\; ,
\end {array}
\label{e5.8}
\end {equation}
\begin {equation}
\begin {array}{l} \displaystyle
  \bar{u}(p_f,n_f) Q v(p_i,n_i)
                        \\[0.5cm] \displaystyle
= { Tr \left[ Q
(- m_i \pm \hat{p}_i \hat{n}_i
  + \hat{p}_i \mp m_i \hat{n}_i )
  ( 1 \mp \gamma _5 ) \hat{q}
( m_f \mp \hat{p}_f \hat{n}_f
  + \hat{p}_f \pm m_f \hat{n}_f ) \right] \over
8 \sqrt{ \left[ ( q p_i ) \mp  m_i ( q n_i ) \right]
\left[ ( q p_f ) \pm  m_f ( q n_f ) \right] } }
\;\; ,
\end {array}
\label{e5.9}
\end {equation}
\begin {equation}
\begin {array}{l} \displaystyle
  \bar{v}(p_f,n_f) Q u(p_i,n_i)
                        \\[0.5cm] \displaystyle
= { Tr \left[ Q
( m_i \pm \hat{p}_i \hat{n}_i
  + \hat{p}_i \pm m_i \hat{n}_i )
  ( 1 \mp \gamma_5 ) \hat{q}
( - m_f \mp \hat{p}_f \hat{n}_f
  + \hat{p}_f \mp m_f \hat{n}_f ) \right] \over
8 \sqrt{ \left[ ( q p_i ) \pm  m_i ( q n_i ) \right]
\left[ ( q p_f ) \mp  m_f ( q n_f ) \right] } }
\;\; ,
\end {array}
\label{e5.10}
\end {equation}
\begin {equation}
\begin {array}{l} \displaystyle
  \bar{v}(p_f,n_f) Q v(p_i,n_i)
                        \\[0.5cm] \displaystyle
= { Tr \left[ Q
(- m_i \pm \hat{p}_i \hat{n}_i
  + \hat{p}_i \mp m_i \hat{n}_i )
  ( 1 \mp \gamma_5 ) \hat{q}
( - m_f \mp \hat{p}_f \hat{n}_f
  + \hat{p}_f \mp m_f \hat{n}_f ) \right] \over
8 \sqrt{ \left[ ( q p_i ) \mp  m_i ( q n_i ) \right]
\left[ ( q p_f ) \mp  m_f ( q n_f ) \right] } }
\;\; .
\end {array}
\label{e5.11}
\end {equation}
    As noted above, $Q$ is the matrix operator which characterizes
the interaction.   It is a  linear combination of  the products of
the   Dirac   $\gamma$-matrices   (or   their   contractions  with
4-vectors).

    If in the course of numerical calculations the denominators in
(\ref{e5.1})~--~(\ref{e5.11})  become  zero  for  some  values  of
vectors
$p_i, \, n_i$
    and
$p_f, \,  n_f$,
    then it suffice  to change the  values of arbitrary  4-vectors
$q$ or $r, \, l$  contained in these formulae (simultaneously  for
all the lines of the diagrams being considered).

    To simplify the calculations one can substitute the vectors of
the problem instead of arbitrary vectors $q$ or $r, \, l$ but this
may cause an ambiguity of the type
$
\displaystyle
{0 \over 0}
\,
$
    which can not be resolved.

    The listed  formulae are  simple enough  and may  be used  for
creating the  computer programs  for automatic  calculation of the
amplitudes.  Besides the presence of the multipliers
$
\displaystyle
( 1 \pm \gamma_5 )
$
    in all the formulae allows us to use the formulae of the Fierz
transformations (see,  e.g.   \cite{g2}), if  it is  necessary, to
simplify the calculations:
\begin {equation}
\displaystyle
 \left[ ( 1 \pm \gamma_5 ) \gamma_{\mu} \right]_{ij}
 \left[ ( 1 \mp \gamma_5 ) \gamma^{\mu} \right]_{kl}
= 2 [ 1 \pm \gamma_5 ]_{il} [ 1 \mp \gamma_5 ]_{kj}
\label{e5.12}
\end {equation}
\begin {equation}
\displaystyle
 \left[ (1 \pm \gamma_5 ) \gamma_{\mu} \right]_{ij}
 \left[ (1 \pm \gamma_5 ) \gamma^{\mu} \right]_{kl}
= - \left[ ( 1 \pm \gamma_5 ) \gamma_{\mu} \right]_{il}
    \left[ ( 1 \pm \gamma_5 ) \gamma^{\mu} \right]_{kj}
\label{e5.13}
\end {equation}
    ($i$, $j$, $k$, $l$ are  indices that label the components  of
$4 \times 4$-matrices).


    \section {Methods of the  calculation of the amplitudes  based
on  multiplication  of  $\gamma$-matrices  and bispinors which are
written in the matrix form and methods based on transformation  of
bispinors}

    Throughout  this Section we use the chiral representation
of the $\gamma$-matrices:
$$
\gamma^0 =
\left(
\begin{array}{cc}
O & I \\ I & O
\end{array}
\right)
,
\gamma^k =
\left(
\begin{array}{cc}
O & \sigma^k \\ -\sigma^k & O
\end{array}
\right)
,
\gamma_5 = {\it i} \gamma^0 \gamma^1 \gamma^2 \gamma^3 =
\left(
\begin{array}{cc}
-I & O \\ O & I
\end{array}
\right)
, \\
$$
$$
\sigma^1 =
\left(
\begin{array}{cc}
0 & 1 \\ 1 & 0
\end{array}
\right)
,
\sigma^2 =
\left(\begin{array}{cc}
0 & -{\it i} \\ {\it i} & 0
\end{array}
\right)
,
\sigma^3 =
\left(
\begin{array}{cc}
1 & 0 \\ 0 & -1
\end{array}
\right)
,
I = \left( \begin{array}{cc} 1&0 \\ 0&1 \end{array} \right),
O = \left( \begin{array}{cc} 0&0 \\ 0&0 \end{array} \right).
$$
    In this representation the  projection operator for a  massive
particle has the form:
\begin {equation}
\begin {array}{l}   \displaystyle
{\cal P}(p,n) = { 1 \over 2 } ( \hat{p} + m )
                ( 1+ \gamma_5 \hat{n} ) = { 1 \over 2 } \cdot
                        \\[0.5cm] \displaystyle
\left( \begin{array}{llll}
m + [pn]_{0z} - {\it i} [pn]_{xy}   &
p_{-} n_{\bot}^{\ast} - p_{\bot}^{\ast} n_{-}   &
p_{-} - m n_{-}   &
-p_{\bot}^{\ast} + m n_{\bot}^{\ast}   \\
p_{+} n_{\bot} - p_{\bot} n_{+}   &
m - [pn]_{0z} + {\it i} [pn]_{xy}   &
-p_{\bot} + m n_{\bot}   &
p_{+} - m n_{+}    \\
p_{+} + m n_{+}   &
p_{\bot}^{\ast} + m n_{\bot}^{\ast}   &
m + [pn]_{0z} + {\it i} [pn]_{xy}   &
p_{+} n_{\bot}^{\ast} - p_{\bot}^{\ast} n_{+}   \\
p_{\bot} + m n_{\bot}   &
p_{-} + m n_{-}   &
p_{-} n_{\bot} - p_{\bot} n_{-}   &
m - [pn]_{0z} - {\it i} [pn]_{xy}
\end{array}\right)
\end {array}
\label{e6.1}
\end {equation}
    where for any 4-vectors $a$ and $b$
$$
a_{\pm} = a_0 \pm a_z
\; , \;\;
a_{\bot} = a_x + {\it i} a_y
\; , \;\;
[ab]_{0z} = a_0 b_z - a_z b_0
\; , \;\;
[ab]_{xy} = a_x b_y - a_y b_x
\; ,
$$
$$
\displaystyle
p_0 = E = \sqrt { m^2 + {\vec{p} \, }^2 }
$$

     The projection operator for a massive antiparticle:
\begin {equation}
\begin {array}{l} \displaystyle
{\cal P}_{a}(p,n) = { 1 \over 2 } ( \hat{p} - m )
                   ( 1 + \gamma_5 \hat{n} ) = { 1 \over 2 } \cdot
                        \\[0.5cm] \displaystyle
\left( \begin{array}{llll}
-m + [pn]_{0z} - {\it i} [pn]_{xy}   &
p_{-} n_{\bot}^{\ast} - p_{\bot}^{\ast} n_{-}   &
p_{-} + m n_{-}   &
-p_{\bot}^{\ast} - m n_{\bot}^{\ast}   \\
p_{+} n_{\bot} - p_{\bot} n_{+}   &
-m - [pn]_{0z} + {\it i} [pn]_{xy}   &
-p_{\bot} - m n_{\bot}   &
p_{+} + m n_{+}    \\
p_{+} - m n_{+}   &
p_{\bot}^{\ast} - m n_{\bot}^{\ast}   &
-m + [pn]_{0z} + {\it i} [pn]_{xy}   &
p_{+} n_{\bot}^{\ast} - p_{\bot}^{\ast} n_{+}   \\
p_{\bot} - m n_{\bot}   &
p_{-} - m n_{-}   &
p_{-} n_{\bot} - p_{\bot} n_{-}   &
-m - [pn]_{0z} - {\it i} [pn]_{xy}
\end{array}\right)
\end {array}
\label{e6.2}
\end {equation}

    As  noted  above,  the  calculations  of  the  amplitudes   by
multiplication  of  $\gamma$-matrices  and  bispinors  which   are
written  in  the  matrix  form  are  reduced to the general scheme
(\ref{e2.2}).  Let us illustrate this by several examples.
\begin{description}
\item[1.1.]
    Let us consider
\begin {equation}
\begin{array}{l} \displaystyle
u(p,n)=  {  {\cal P}(p,n) \over
   \sqrt { Tr \left[ {\cal P}(p,n) {\cal P}(r,l) \right] } }
 \left( \begin{array}{l} 1 \\ 0 \\ 1 \\ 0
 \end{array} \right)
\\           \\  \displaystyle
 = { 1 \over  2 \sqrt{ ( p_0 + m ) ( 1 + n_z ) - p_z n_0 } }
 \left( \begin{array}{l}
   p_{-} + m (1 - n_{-}) + [pn]_{0z} - {\it i}[pn]_{xy}  \\
   (m + p_{+}) n_{\bot} - p_{\bot}(1 + n_{+})            \\
   p_{+} + m(1 + n_{+}) + [pn]_{0z} + {\it i}[pn]_{xy}   \\
   (m + p_{-}) n_{\bot} + p_{\bot}(1 - n_{-})
 \end{array} \right)
 \;\; ,
\end{array}
\label{e6.3}
\end {equation}
    where
$
r^{\mu} = (1,0,0,0)
\;\; , \;\;\;
l^{\mu} = (0,0,0,1)
\;\; .
$
\begin {equation}
\begin{array}{l} \displaystyle
u'(p,n)
 = {  {\cal P}(p,n) \over
   \sqrt { Tr \left[ {\cal P}(p,n) {\cal P}(r,l') \right] } }
 \left( \begin{array}{l} 0 \\ 1 \\ 0 \\ 1
 \end{array} \right)
       \\    \\  \displaystyle
= { 1 \over 2 \sqrt{ ( p_0 + m ) ( 1 - n_z ) + p_z n_0 } }
 \left( \begin{array}{l}
   (m + p_{-}) n_{\bot}^{\ast} - p_{\bot}^{\ast}(1 + n_{-})   \\
   p_{+} + m (1 - n_{+}) - [pn]_{0z} + {\it i}[pn]_{xy}       \\
   (m + p_{+}) n_{\bot}^{\ast} + p_{\bot}^{\ast}(1 - n_{+})   \\
   p_{-} + m (1 + n_{-}) - [pn]_{0z} - {\it i}[pn]_{xy}
 \end{array} \right)
 \;\; ,
\end{array}
\label{e6.4}
\end {equation}
    where
$
r^{\mu} = (1,0,0,0)
\;\; , \;\;\;
{l'}^{\mu} = (0,0,0,-1)
\;\; .
$
    Note that
\begin{equation}
\displaystyle
u'(p,n) = e^{ {\it i} \varphi(p,n) } u(p,n)
\;\; ,
\label{e6.5}
\end{equation}
    where
\begin{equation}
\displaystyle
e^{ {\it i} \varphi(p,n) } =
{ (p_0 + m) n_{\bot}^{\ast} - p_{\bot}^{\ast} n_0   \over
\sqrt{ ( p_0 + m ) ( 1 - n_z ) + p_z n_0 }
\sqrt{ ( p_0 + m ) ( 1 + n_z ) - p_z n_0 } }
\;\; .
\label{e6.6}
\end{equation}

    Similarly:
\begin {equation}
\begin{array}{l} \displaystyle
v(p,n)
 = { {\cal P}_{a}(p,n) \over
   \sqrt{ Tr \left[ {\cal P}_{a}(p,n) {\cal P}(r,l) \right] } }
 \left( \begin{array}{l} 1 \\ 0 \\ 1 \\ 0
 \end{array} \right)
         \\  \\  \displaystyle
= { 1 \over  2 \sqrt{ (p_0 - m) (1 + n_z) - p_z n_0 } }
 \left( \begin{array}{l}
   p_{-} - m (1 - n_{-}) + [pn]_{0z} - {\it i}[pn]_{xy}  \\
   (-m + p_{+}) n_{\bot} - p_{\bot}(1 + n_{+})            \\
   p_{+} - m(1 + n_{+}) + [pn]_{0z} + {\it i}[pn]_{xy}   \\
   (-m + p_{-}) n_{\bot} + p_{\bot}(1 - n_{-})
 \end{array} \right)
\;\; ,
\end{array}
\label{e6.7}
\end {equation}
\begin {equation}
\begin{array}{l} \displaystyle
v'(p,n) = { {\cal P}_a (p,n) \over
   \sqrt { Tr \left[ {\cal P}_a (p,n) {\cal P}(r,l') \right] } }
 \left( \begin{array}{l} 0 \\ 1 \\ 0 \\ 1
 \end{array} \right)
  \\         \\  \displaystyle
 = { 1 \over 2 \sqrt{ (p_0 - m) (1 - n_z) + p_z n_0 } }
 \left( \begin{array}{l}
   (-m + p_{-}) n_{\bot}^{\ast} - p_{\bot}^{\ast}(1 + n_{-})   \\
   p_{+} - m (1 - n_{+}) - [pn]_{0z} + {\it i}[pn]_{xy}       \\
   (-m + p_{+}) n_{\bot}^{\ast} + p_{\bot}^{\ast}(1 - n_{+})   \\
   p_{-} - m (1 + n_{-}) - [pn]_{0z} - {\it i}[pn]_{xy}
 \end{array} \right)
\;\; .
\end{array}
\label{e6.8}
\end {equation}
    Note that
\begin{equation}
\displaystyle
v'(p,n) = e^{ {\it i} \phi(p,n) } v(p,n)
\;\; ,
\label{e6.9}
\end{equation}
    where
\begin{equation}
\displaystyle
e^{ {\it i} \phi(p,n) } =
{ (p_0 - m) n_{\bot}^{\ast} - p_{\bot}^{\ast} n_0   \over
\sqrt{ (p_0 - m) (1 - n_z) + p_z n_0 }
\sqrt{ (p_0 - m) (1 + n_z) - p_z n_0 } }
\;\; .
\label{e6.10}
\end{equation}

    Obviously  the   bispinors  (\ref{e6.3})~--~(\ref{e6.4})   for
particles  and  the  bispinors  (\ref{e6.7})~--~(\ref{e6.8})   for
antiparticles satisfy as the  Dirac equation so also  the equation
for the vector $n$ determining  the axis of the spin  projections,
since  when  constructing  the  bispinors  there  were  used   the
corresponding  projection  operators  (\ref{e6.1}),  (\ref{e6.2}).
Besides one  can easily  verify that  these bispinors  satisfy the
normalization conditions
$$
\begin{array}{c} \displaystyle
   \bar{u}(p,n) u(p,n)
 = \bar{u}'(p,n) u'(p,n) = 2 m
\; ,
 \\    \\   \displaystyle
   \bar{v}(p,n) v(p,n)
 = \bar{v}'(p,n) v'(p,n) = - 2 m
\; .
\end{array}
$$

    It follows from (\ref{e6.3}):
\begin{equation}
\begin{array}{l} \displaystyle
\bar{u}(p_f,n_f) Q u(p_i,n_i) =
Tr \left[ Q u(p_i,n_i) \bar{u} (p_f,n_f) \right]
 \\    \\   \displaystyle
= Tr \left[ Q { {\cal P}(p_i,n_i) \over
\sqrt{ Tr \left[ {\cal P}(p_i,n_i) {\cal P}(r,l) \right] }}
\left( \begin{array}{llll}
1   &   0    &   1   &   0   \\
0   &   0    &   0   &   0   \\
1   &   0    &   1   &   0   \\
0   &   0    &   0   &   0
\end{array}\right)
   { {\cal P}(p_f,n_f) \over
\sqrt{ Tr \left[ {\cal P}(p_f,n_f) {\cal P}(r,l) \right] }}
   \right]
    \\  \\  \displaystyle
= { Tr \left[ Q {\cal P}(p_i,n_i) {\cal P}(r,l)
                {\cal P}(p_f,n_f) \right] \over
\sqrt{ Tr \left[ {\cal P}(p_i,n_i) {\cal P}(r,l) \right]
       Tr \left[ {\cal P}(p_f,n_f) {\cal P}(r,l) \right] } }
\end{array}
\label{e6.11}
\end{equation}
    i.e. the calculation of the amplitude in this case reduces  to
the general scheme (\ref{e2.2}) if
$$
\displaystyle
Z = {\cal P}(r,l) = { 1 \over 2 } ( 1 + \gamma ^0 )
( 1 - \gamma_5 \gamma^3 )
\;\; .
$$

     Similarly
\begin{equation}
\begin{array}{l} \displaystyle
\bar{u}'(p_f,n_f) Q u'(p_i,n_i)
 \\    \\   \displaystyle
= Tr \left[ Q { {\cal P}(p_i,n_i) \over
\sqrt{ Tr \left[ {\cal P}(p_i,n_i) {\cal P}(r,l')
\right] }}
\left( \begin{array}{llll}
0   &   0    &   0   &   0   \\
0   &   1    &   0   &   1   \\
0   &   0    &   0   &   0   \\
0   &   1    &   0   &   1
\end{array}\right)
   { {\cal P}(p_f,n_f) \over
\sqrt{ Tr \left[ {\cal P}(p_f,n_f) {\cal P}(r,l')
\right] }}
   \right]
    \\  \\  \displaystyle
= { Tr \left[ Q {\cal P}(p_i,n_i) {\cal P}(r,l')
           {\cal P}(p_f,n_f) \right] \over
\sqrt{ Tr \left[ {\cal P}(p_i,n_i) {\cal P}(r,l')
\right]
       Tr \left[ {\cal P}(p_f,n_f) {\cal P}(r,l')
       \right] } }
\;\; .
\end{array}
\label{e6.12}
\end {equation}
    Here
$$
\displaystyle
Z =  {\cal P}(r,l') = { 1 \over 2 } ( 1 + \gamma ^0 )
( 1 + \gamma_5 \gamma^3 )
\;\; .
$$
    And,  as  it  follows  from  (\ref{e6.5})~--~(\ref{e6.6}), the
expressions  obtained  for  the  amplitudes  differ  by  the phase
factor.

    Similarly, for the amplitude
$$
\displaystyle
\bar{u}(p_f,n_f) Q u'(p_i,n_i)
$$
    we have
$$
\displaystyle
Z = {\cal P}(r,l) \cdot e^{ {\it i} \varphi(p_i,n_i) }
$$
    and so on.

    Note that until the present time the bispinors of the  general
form  (\ref{e6.3})~--~(\ref{e6.4}),   (\ref{e6.7})~--~(\ref{e6.8})
have  not  been  used  for  the  calculation  of the amplitudes of
processes, however their special forms described below are  widely
used.

\item[1.2.]
    Let    us    consider    the    form    of    the    bispinors
(\ref{e6.3})~--~(\ref{e6.4}), (\ref{e6.7})~--~(\ref{e6.8}) in  the
situation   when   the   polarization   state   of  particles  and
antiparticles is the helicity, i.e.
$$
\displaystyle
n^{\mu}(p) = { \lambda \over m } \left( \left| \vec{p} \right| ,
  { p_0 \over \left| \vec{p} \right| } \vec{p} \right)
\;\; , \;\;\;
\lambda = \pm 1
\;\; .
$$
    In this case the bispinors take the form
\begin {equation}
\begin{array}{l} \displaystyle
u(p,+) = { 1 \over 2 \sqrt{ ( p_0 + m)  \left| \vec p \right|
                  \left( \left| \vec p \right| + p_z \right) } }
 \left( \begin{array}{l}
        \left( p_0 + m - \left| \vec p \right| \right)
                  \left( \left| \vec p \right| + p_z \right) \\
        \left( p_0 + m - \left| \vec p \right| \right)
                  \quad p_{\bot} \\
        \left( p_0 + m + \left| \vec p \right| \right)
                  \left( \left| \vec p \right| + p_z \right) \\
        \left( p_0 + m + \left| \vec p \right| \right)
                  \quad p_{\bot}
 \end{array} \right)
                        \\  \\  \displaystyle
= { 1 \over \sqrt{ 2  \left| \vec p \right|
                  \left( \left| \vec p \right| + p_z \right) } }
\left( \begin{array}{l}
        \sqrt{ p_0 -  \left| \vec p \right| }
               \left( \left| \vec p \right| + p_z \right) \\
        \sqrt{ p_0 -  \left| \vec p \right| }
               \quad p_{\bot} \\
        \sqrt{ p_0 +  \left| \vec p \right| }
               \left( \left| \vec p \right| + p_z \right) \\
        \sqrt{ p_0 +  \left| \vec p \right| }
               \quad p_{\bot}
 \end{array} \right)
\;\; .
\end{array}
\label{e6.13}
\end {equation}
    In (\ref{e6.13}) we used the identity:
$$
\displaystyle
{ p_0 + m \pm  \left| \vec p \right|  \over
                           \sqrt{ 2 ( p_0 + m ) } }
= \sqrt{ p_0 \pm \left| \vec p \right| }
\;\; .
$$
    Further
\begin {equation}
\begin{array}{l} \displaystyle
u(p,-) = { 1 \over 2 \sqrt{ ( p_0 + m ) \left| \vec p \right|
                 \left( \left| \vec p \right| - p_z \right) } }
 \left( \begin{array}{l}
       \left( p_0 + m + \left| \vec p \right| \right)
                 \left( \left| \vec p \right| - p_z \right) \\
       \left( p_0 + m + \left| \vec p \right| \right)
                 \quad \left( - p_{\bot} \right) \\
       \left( p_0 + m - \left| \vec p \right| \right)
                 \left( \left| \vec p \right| - p_z \right) \\
       \left( p_0 + m - \left| \vec p \right| \right)
                 \quad \left( - p_{\bot} \right)
 \end{array} \right)
                       \\  \\  \displaystyle
= { 1 \over \sqrt{ 2  \left| \vec p \right|
               \left( \left| \vec p \right| - p_z \right) } }
 \left( \begin{array}{l}
       \sqrt{ p_0 +  \left| \vec p \right| }
              \left( \left| \vec p \right| - p_z \right) \\
       \sqrt{ p_0 +  \left| \vec p \right| }
              \quad \left( - p_{\bot} \right) \\
       \sqrt{ p_0 -  \left| \vec p \right| }
              \left( \left| \vec p \right| - p_z \right) \\
       \sqrt{ p_0 -  \left| \vec p \right| }
              \quad \left( - p_{\bot} \right)
 \end{array} \right)
\;\; ,
\end{array}
\label{e6.14}
\end {equation}
\begin {equation}
\begin{array}{l} \displaystyle
u'(p,+)
= { 1 \over 2 \sqrt{ ( p_0 + m) \left| \vec p \right|
                   \left( \left| \vec p \right| - p_z \right) } }
\left( \begin{array}{l}
    \left( p_0 + m - \left| \vec p \right| \right)
              \quad p_{\bot}^{\ast}   \\
    \left( p_0 + m - \left| \vec p \right| \right)
              \left( \left| \vec p \right| - p_z \right) \\
    \left( p_0 + m + \left| \vec p \right| \right)
              \quad  p_{\bot}^{\ast}   \\
    \left( p_0 + m + \left| \vec p \right| \right)
              \left( \left| \vec p \right| - p_z \right)
 \end{array} \right)
    \\  \\  \displaystyle
= { 1 \over \sqrt{ 2  \left| \vec p \right|
   \left( \left| \vec p \right| - p_z \right) } }
 \left( \begin{array}{l}
      \sqrt{ p_0 - \left| \vec p \right| }
             \quad  p_{\bot}^{\ast}   \\
      \sqrt{ p_0 -  \left| \vec p \right| }
             \left( \left| \vec p \right| - p_z \right) \\
      \sqrt{ p_0 +  \left| \vec p \right| }
             \quad  p_{\bot}^{\ast}   \\
      \sqrt{ p_0 +  \left| \vec p \right| }
             \left( \left| \vec p \right| - p_z \right)
 \end{array} \right)
\;\; ,
\end{array}
\label{e6.15}
\end {equation}
\begin {equation}
\begin{array}{l} \displaystyle
u'(p,-)
= { 1 \over 2 \sqrt{ ( p_0 + m)  \left| \vec p \right|
   \left( \left| \vec p \right| + p_z \right) } }
\left( \begin{array}{l}
    \left( p_0 + m + \left| \vec p \right| \right)
              \left( - p_{\bot}^{\ast}  \right) \\
    \left( p_0 + m + \left| \vec p \right| \right)
              \left( \left| \vec p \right| + p_z \right) \\
    \left( p_0 + m - \left| \vec p \right| \right)
              \left( - p_{\bot}^{\ast}  \right) \\
    \left( p_0 + m - \left| \vec p \right| \right)
              \left( \left| \vec p \right| + p_z \right)
 \end{array} \right)
    \\  \\  \displaystyle
= { 1 \over \sqrt{ 2  \left|  \vec p \right|
   \left( \left|  \vec p \right| + p_z \right) } }
 \left( \begin{array}{l}
     \sqrt{ p_0 +  \left| \vec p \right| }
            \left( - p_{\bot}^{\ast} \right) \\
     \sqrt{ p_0 +  \left| \vec p \right| }
            \left( \left| \vec p \right| + p_z \right) \\
     \sqrt{ p_0 -  \left| \vec p \right| }
            \left( - p_{\bot}^{\ast} \right) \\
     \sqrt{ p_0 -  \left| \vec p \right| }
            \left( \left| \vec p \right| + p_z \right)
 \end{array} \right)
\;\; .
\end{array}
\label{e6.16}
\end {equation}
    As this takes place we have
$$
\displaystyle
u'(p,\pm) = \pm u(p,\pm) \cdot
{ p_{\bot}^{\ast}  \over  \sqrt{ p_x^2 + p_y^2 } }
 = \pm u(p,\pm) \cdot
{ p_{\bot}^{\ast}  \over \left| p_{\bot} \right| }
\;\; .
$$
    Further
\begin {equation}
\begin{array}{l} \displaystyle
v(p,+) = { 1 \over 2 \sqrt{ ( p_0 - m) \left| \vec p \right|
   \left( \left| \vec p \right| - p_z \right) } }
 \left( \begin{array}{l}
      \left( p_0 - m + \left| \vec p \right| \right)
                \left( \left| \vec p \right| - p_z \right) \\
      \left( p_0 - m + \left| \vec p \right| \right)
                \quad \left( - p_{\bot} \right) \\
      \left( p_0 - m - \left| \vec p \right| \right)
                \left( \left| \vec p \right| - p_z \right) \\
      \left( p_0 - m - \left| \vec p \right| \right)
                \quad \left( - p_{\bot} \right)
 \end{array} \right)
    \\  \\  \displaystyle
= { 1 \over \sqrt{ 2  \left| \vec p \right|
   \left( \left| \vec p \right| - p_z \right) } }
 \left( \begin{array}{l}
        \sqrt{ p_0 + \left| \vec p \right|}
              \left( \left| \vec p \right| - p_z \right) \\
        \sqrt{ p_0 + \left| \vec p \right| }
               \quad \left( - p_{\bot} \right) \\
     -  \sqrt{ p_0 -  \left| \vec p \right| }
               \left( \left| \vec p \right| - p_z \right) \\
     -  \sqrt{ p_0 -  \left| \vec p \right| }
               \quad \left( - p_{\bot} \right)
 \end{array} \right)
\;\; .
\end{array}
\label{e6.17}
\end {equation}
    In (\ref{e6.17}) we used the identity:
$$
\displaystyle
{ p_0 - m \pm  \left| \vec p \right| \over
                                     \sqrt{ 2 ( p_0 - m ) } }
= \pm \sqrt{ p_0 \pm  \left| \vec p \right| }
\;\; .
$$
\begin {equation}
\begin{array}{l} \displaystyle
v(p,-) = { 1 \over 2 \sqrt{ ( p_0 - m)  \left| \vec p \right|
   \left( \left| \vec p \right| + p_z \right) } }
\left( \begin{array}{l}
      \left( p_0 - m - \left| \vec p \right| \right)
                \left( \left| \vec p \right| + p_z \right) \\
      \left( p_0 - m - \left| \vec p \right| \right)
                \quad p_{\bot} \\
      \left( p_0 - m + \left| \vec p \right| \right)
                \left( \left| \vec p \right| + p_z \right) \\
      \left( p_0 - m + \left| \vec p \right| \right)
                \quad p_{\bot}
 \end{array} \right)
    \\  \\  \displaystyle
= { 1 \over \sqrt{ 2  \left| \vec p \right|
   \left( \left| \vec p \right| + p_z \right) } }
 \left( \begin{array}{l}
      - \sqrt{ p_0 -  \left| \vec p \right| }
               \left( \left| \vec p \right| + p_z \right) \\
      - \sqrt{ p_0 -  \left| \vec p \right| }
               \quad p_{\bot} \\
        \sqrt{ p_0 +  \left| \vec p \right| }
               \left( \left| \vec p \right| + p_z \right) \\
        \sqrt{ p_0 +  \left| \vec p \right| }
               \quad p_{\bot}
 \end{array} \right)
\;\; ,
\end{array}
\label{e6.18}
\end {equation}
\begin {equation}
\begin{array}{l} \displaystyle
v'(p,+)
= { 1 \over 2 \sqrt{ ( p_0 - m)  \left| \vec p \right|
   \left( \left| \vec p \right| + p_z \right) } }
\left( \begin{array}{l}
    \left( p_0 - m + \left| \vec p \right| \right)
              \left( - p_{\bot}^{\ast}  \right) \\
    \left( p_0 - m + \left| \vec p \right| \right)
              \left( \left| \vec p \right| + p_z \right) \\
    \left( p_0 - m - \left| \vec p \right| \right)
              \left( - p_{\bot}^{\ast}  \right) \\
    \left( p_0 - m - \left| \vec p \right| \right)
              \left( \left| \vec p \right| + p_z \right)
 \end{array} \right)
                     \\  \\  \displaystyle
= { 1 \over \sqrt{ 2  \left|  \vec p \right|
                  \left( \left|  \vec p \right| + p_z \right) } }
 \left( \begin{array}{l}
        \sqrt{ p_0 +  \left| \vec p \right|}
               \left( - p_{\bot}^{\ast}  \right) \\
        \sqrt{ p_0 +  \left| \vec p \right| }
               \left( \left| \vec p \right| + p_z \right) \\
      - \sqrt{ p_0 -  \left| \vec p \right| }
               \left( - p_{\bot}^{\ast} \right) \\
      - \sqrt{ p_0 -  \left| \vec p \right| }
               \left( \left| \vec p \right| + p_z \right)
 \end{array} \right)
\;\; ,
\end{array}
\label{e6.19}
\end {equation}
\begin {equation}
\begin{array}{l} \displaystyle
v'(p,-)
= { 1 \over 2 \sqrt{ ( p_0 - m)  \left| \vec p \right|
                    \left( \left| \vec p \right| - p_z \right) } }
\left( \begin{array}{l}
    \left( p_0 - m - \left| \vec p \right| \right)
              \quad  p_{\bot}^{\ast}   \\
    \left( p_0 - m - \left| \vec p \right| \right)
              \left( \left| \vec p \right| - p_z \right) \\
    \left( p_0 - m + \left| \vec p \right| \right)
              \quad  p_{\bot}^{\ast}   \\
    \left( p_0 - m + \left| \vec p \right| \right)
              \left( \left| \vec p \right| - p_z \right)
 \end{array} \right)
                      \\  \\  \displaystyle
= { 1 \over \sqrt{ 2  \left| \vec p \right|
   \left( \left|  \vec p \right| - p_z \right) } }
 \left( \begin{array}{l}
    - \sqrt{ p_0 - \left| \vec p \right|}
             \quad  p_{\bot}^{\ast}   \\
    - \sqrt{ p_0 -  \left| \vec p \right| }
             \left( \left| \vec p \right| - p_z \right) \\
      \sqrt{ p_0 +  \left| \vec p \right| }
             \quad  p_{\bot}^{\ast}   \\
      \sqrt{ p_0 +  \left| \vec p \right| }
             \left( \left| \vec p \right| - p_z \right)
 \end{array} \right)
\;\; .
\end{array}
\label{e6.20}
\end {equation}
    As this takes place we obtain
$$
\displaystyle
v'(p,\pm) = \mp v(p,\pm) \cdot
 { p_{\bot}^{\ast}  \over  \sqrt{ p_x^2 + p_y^2 } }
                  = \mp v(p,\pm) \cdot
 { p_{\bot}^{\ast}  \over  \left| p_{\bot} \right| }
\;\; .
$$
    (The signs $\pm$  in notation of  the bispinors correspond  to
the helicity of particles or antiparticles.)

    The   bispinors   (\ref{e6.13})~--~(\ref{e6.20})   have   been
considered in a number of papers.

    In particular, in \cite{r6.1.2-1} the bispinors (\ref{e6.13}),
(\ref{e6.14}) were considered.

    In \cite{r6.1.2-2} the bispinors (\ref{e6.13}), (\ref{e6.16})
were considered.

    In  \cite{r6.1.2-3}  it  is  proposed  to  use  the  bispinors
(\ref{e6.13}),  (\ref{e6.16}),  (\ref{e6.18}),  (\ref{e6.19})  [in
this  paper  the  bispinors  (\ref{e6.18}), (\ref{e6.19}) have the
opposite sign].

    In  \cite{r6.1.2-4}  it  is  proposed  to  use  the  bispinors
(\ref{e6.13})~--~(\ref{e6.20})  [in   this  paper   the  bispinors
(\ref{e6.17}), (\ref{e6.19}) have the opposite sign].

    In \cite{r6.1.2-5} the bispinors (\ref{e6.13}), (\ref{e6.16}),
(\ref{e6.18}), (\ref{e6.19}) are used [in this paper the  bispinor
(\ref{e6.19}) has the opposite sign].

    In \cite{r6.1.2-6} there are presented the expressions for the
amplitudes  that  can  be  obtained  by  means  of  the   formulae
(\ref{e6.14}), (\ref{e6.15}), (\ref{e6.18}), (\ref{e6.19}) written
in the center-of-mass frame of the fermions considered.

    The bispinors  described in  this subsection  are also  widely
used in calculations others authors.

\item[1.3.]
    Let us consider a special form of the bispinors
(\ref{e6.13})~--~(\ref{e6.20}) at
$
\displaystyle
m = 0 \quad \left( \left| \vec p \right| = p_0  \right) :
$
\begin{equation}
\displaystyle
u_{+}(p)= v_{-}(p)={ 1 \over \sqrt{ p_{+} }}
 \left( \begin{array}{c} 0 \\ 0 \\ p_{+} \\ p_{\bot}
 \end{array} \right)
\;\; ,
\label{e6.21}
\end{equation}
\begin{equation}
\displaystyle
u_{-}(p)= v_{+}(p)=
{ 1 \over \sqrt{ p_{-} }}
 \left( \begin{array}{c} p_{-} \\ -p_{\bot} \\ 0 \\ 0
 \end{array} \right)
\;\; ,
\label{e6.22}
\end{equation}
\begin{equation}
\displaystyle
u_{+}'(p) = v_{-}'(p)={ 1 \over \sqrt{ p_{-} }}
 \left( \begin{array}{c} 0 \\ 0 \\ p_{\bot}^{\ast} \\ p_{-}
 \end{array} \right)
\;\; ,
\label{e6.23}
\end{equation}
\begin{equation}
\displaystyle
u_{-}'(p) = v_{+}'(p) =
{ 1 \over \sqrt{ p_{+} } }
 \left( \begin{array}{c} -p_{\bot}^{\ast} \\ p_{+} \\ 0 \\ 0
 \end{array} \right)
\;\; .
\label{e6.24}
\end{equation}

    In  a  number  of  papers  it  has  been  proposed  to use the
bispinors (\ref{e6.21})~--~(\ref{e6.24}).

    In  \cite{r6.1.3-1}  it  was  proposed  to  use  the bispinors
(\ref{e6.21}), (\ref{e6.24}).

    The  method  proposed  in  \cite{r6.1.3-2}  is  reduced to the
application   of   the   bispinors   of  the  form  (\ref{e6.22}),
(\ref{e6.23}) [bispinor (\ref{e6.22}) was used with the additional
factor ${\it i}$].

    In \cite{r6.1.3-3} it was proposed to use the bispinors of the
form (\ref{e6.21}), (\ref{e6.24}) [here the bispinor (\ref{e6.24})
was used with the opposite sign].

    The  method  proposed   in  \cite{r6.1.3-4}  reduces   to  the
application    of    a    special    form    of   the    bispinors
(\ref{e6.21})~--~(\ref{e6.24}) at
$
\displaystyle
p_y = 0
\; .
$

    These methods are widely used in calculations.

\item[1.4.]
    For%
\footnote{
    In the rest frame of a particle (antiparticle) the vector
$n^{\mu}$
    takes the form
$
\displaystyle
{\lambda} (0, 0, 0, 1)
$.
}
$$
\displaystyle
n^{\mu} = { \lambda \over m } (p_z, { p_x p_z \over p_0 + m },
{ p_y p_z \over p_0 + m }, m + { p_z^2  \over p_0 + m } )
\, , \;
\lambda = \pm 1
$$
    the   formulae   (\ref{e6.3}),   (\ref{e6.4}),   (\ref{e6.7}),
(\ref{e6.8}) imply
\begin{equation}
\displaystyle
u(p,+) = { 1 \over \sqrt{ 2 (p_0 + m) }}
 \left(
\begin{array}{c}
p_{-} + m \\ - p_{\bot} \\ p_{+} + m \\ p_{\bot}
\end{array}
 \right)
\; ,
\label{e6.25}
\end{equation}
\begin{equation}
\displaystyle
u'(p,-) = { 1 \over \sqrt{ 2 (p_0 + m) }}
 \left(
\begin{array}{c}
- p_{\bot}^{\ast} \\ p_{+} + m \\  p_{\bot}^{\ast} \\ p_{-} + m
\end{array}
 \right)
\; ,
\label{e6.26}
\end{equation}
\begin{equation}
\displaystyle
v(p,-) = { 1 \over \sqrt{ 2 (p_0 + m) }}
 \left(
\begin{array}{c}
- p_{-} - m \\  p_{\bot} \\ p_{+} + m \\ p_{\bot}
\end{array}
 \right)
\; ,
\label{e6.27}
\end{equation}
\begin{equation}
\displaystyle
v'(p,+) = { 1 \over \sqrt{ 2 (p_0 + m) }}
 \left(
\begin{array}{c}
- p_{\bot}^{\ast} \\ p_{+} + m \\ - p_{\bot}^{\ast} \\ - p_{-} + m
\end{array}
 \right)
\; .
\label{e6.28}
\end{equation}

    The bispinors (\ref{e6.25})~--~(\ref{e6.28}) are considered in
\cite{r6.1.4-1},     \cite{r6.1.4-2},     \cite{r6.1.4-3}      (in
\cite{r6.1.4-2} and \cite{r6.1.4-3} the bispinor (\ref{e6.28}) has
the opposite sign).

    In \cite{r6.1.4-4} the bispinors (\ref{e6.25}),  (\ref{e6.26})
are used for a special case when the particles are moving on  $xy$
plane, i.e.
$
\displaystyle
p^{\mu} = (p_0, p_x, p_y, 0)
$
    and
$
\displaystyle
n^{\mu} = \lambda (0, 0, 0, 1)
\, , \;
\lambda = \pm 1
$
.

\item[1.5.]
For%
\footnote{
    In the rest frame of a particle (antiparticle) the vector
$n^{\mu}$
    takes the form
$
\displaystyle
{\lambda} (0, 0, 0, 1)
$.
}
$$
\displaystyle
n^{\mu} = { \lambda \over m } ( p_0 - { m^2 \over p_0 + p_z },
p_x, p_y, p_z + { m^2 \over p_0 + p_z } )
\, , \;
\lambda = \pm 1
$$
    the   formulae   (\ref{e6.3}),   (\ref{e6.4}),   (\ref{e6.7}),
(\ref{e6.8}) imply
\begin{equation}
\displaystyle
u(p,+) = { 1 \over \sqrt { p_{+} } }
 \left(
\begin{array}{c}
m \\  0 \\  p_{+}  \\  p_{\bot}
\end{array}
 \right)
\; ,
\label{e6.29}
\end{equation}
\begin{equation}
\displaystyle
u'(p,-)  = { 1 \over \sqrt{ p_{+} } }
 \left(
\begin{array}{c}
- p_{\bot}^{\ast} \\  p_{+}  \\  0  \\  m
\end{array}
 \right)
\; ,
\label{e6.30}
\end{equation}
\begin{equation}
\displaystyle
v(p,-) = { 1 \over \sqrt{ p_{+} } }
 \left(
\begin{array}{c}
- m \\ 0 \\ p_{+} \\ p_{\bot}
\end{array}
 \right)
\; ,
\label{e6.31}
\end{equation}
\begin{equation}
\displaystyle
v'(p,+) = { 1 \over \sqrt{ p_{+} } }
 \left(
\begin{array}{c}
- p_{\bot}^{\ast} \\ p_{+} \\ 0 \\ - m
\end{array}
 \right)
\; .
\label{e6.32}
\end{equation}
    The  bispinors  (\ref{e6.29})~--~(\ref{e6.32})  are  used   in
\cite{r6.1.5-1}, \cite{r7.2-1}.  Note  that in the massless  limit
these bispinors take the form (\ref{e6.21}), (\ref{e6.24}).

\item[2.]

    In the papers  \cite{r6.2-1}, \cite{r6.2-2}, \cite{r6.2-3}  it
is  proposed  to  use  the  following  bispinors for massive Dirac
particles (antiparticles), the polarization state of which is  the
helicity:
\begin {equation}
\displaystyle
u''(p,+)
= { 1 \over \sqrt{ 2  \left| \vec p \right|
                      \left| p_{\bot} \right| } }
\left( \begin{array}{l}
        \sqrt{ p_0 -  \left| \vec p \right| }
 \sqrt{ \left( \left| \vec p \right| + p_z \right)
                                     p_{\bot}^{\ast} } \\
        \sqrt{ p_0 -  \left| \vec p \right| }
 \sqrt{ \left( \left| \vec p \right| - p_z \right)  p_{\bot} } \\
        \sqrt{ p_0 +  \left| \vec p \right| }
 \sqrt{ \left( \left| \vec p \right| + p_z \right)
                                     p_{\bot}^{\ast} } \\
        \sqrt{ p_0 +  \left| \vec p \right| }
 \sqrt{ \left( \left| \vec p \right| - p_z \right)  p_{\bot} }
 \end{array} \right)
\;\; ,
\label{e6.33}
\end {equation}
\begin {equation}
\displaystyle
u''(p,-)
= { 1 \over \sqrt{ 2  \left| \vec p \right|
                      \left| p_{\bot} \right| } }
\left( \begin{array}{r}
      - \sqrt{ p_0 +  \left| \vec p \right| }
 \sqrt{ \left( \left| \vec p \right| - p_z \right)
                                     p_{\bot}^{\ast} } \\
        \sqrt{ p_0 +  \left| \vec p \right| }
 \sqrt{ \left( \left| \vec p \right| + p_z \right)  p_{\bot} } \\
      - \sqrt{ p_0 -  \left| \vec p \right| }
 \sqrt{ \left( \left| \vec p \right| - p_z \right)
                                     p_{\bot}^{\ast} } \\
        \sqrt{ p_0 -  \left| \vec p \right| }
 \sqrt{ \left( \left| \vec p \right| + p_z \right)  p_{\bot} }
 \end{array} \right)
\;\; ,
\label{e6.34}
\end {equation}
\begin {equation}
\displaystyle
v''(p,+)
= { 1 \over \sqrt{ 2  \left| \vec p \right|
                      \left| p_{\bot} \right| } }
\left( \begin{array}{r}
        \sqrt{ p_0 +  \left| \vec p \right| }
 \sqrt{ \left( \left| \vec p \right| - p_z \right)
                                     p_{\bot}^{\ast} } \\
      - \sqrt{ p_0 +  \left| \vec p \right| }
 \sqrt{ \left( \left| \vec p \right| + p_z \right)  p_{\bot} } \\
      - \sqrt{ p_0 -  \left| \vec p \right| }
 \sqrt{ \left( \left| \vec p \right| - p_z \right)
                                     p_{\bot}^{\ast} } \\
        \sqrt{ p_0 -  \left| \vec p \right| }
 \sqrt{ \left( \left| \vec p \right| + p_z \right)  p_{\bot} }
 \end{array} \right)
\;\; ,
\label{e6.35}
\end {equation}
\begin {equation}
\displaystyle
v''(p,-)
= { 1 \over \sqrt{ 2  \left| \vec p \right|
                      \left| p_{\bot} \right| } }
\left( \begin{array}{r}
        \sqrt{ p_0 -  \left| \vec p \right| }
 \sqrt{ \left( \left| \vec p \right| + p_z \right)
                                     p_{\bot}^{\ast} } \\
        \sqrt{ p_0 -  \left| \vec p \right| }
 \sqrt{ \left( \left| \vec p \right| - p_z \right)  p_{\bot} } \\
      - \sqrt{ p_0 +  \left| \vec p \right| }
 \sqrt{ \left( \left| \vec p \right| + p_z \right)
                                     p_{\bot}^{\ast} } \\
      - \sqrt{ p_0 +  \left| \vec p \right| }
 \sqrt{ \left( \left| \vec p \right| - p_z \right)  p_{\bot} }
 \end{array} \right)
\label{e6.36}
\end {equation}
    (In  \cite{r6.2-1}  the  bispinors  for antiparticles have the
opposite   sign,   and   in   \cite{r6.2-3}   the   bispinors  for
antiparticles have an additional factor ${\it i}$).

    As      this      takes      place      we      have      [see
(\ref{e6.13})~--~(\ref{e6.14}), (\ref{e6.17})~--~(\ref{e6.18})]:
$$
\displaystyle
u''(p,\pm) = \pm u(p,\pm) \cdot
 \sqrt{ p_{\bot}^{\ast}  \over \left| p_{\bot} \right| }
\;\; , \;\;
v''(p,\pm) = \pm v(p,\pm) \cdot
 \sqrt{ p_{\bot}^{\ast}  \over \left| p_{\bot} \right| }
\;\; .
$$
    Thus     if     we     use     the     bispinors    in    form
(\ref{e6.33})~--~(\ref{e6.36})   then   the   calculation  of  the
amplitudes  of  processes  is   reduced  to  the  general   scheme
(\ref{e2.2}).  For example for the amplitude
$$
\displaystyle
\bar{u}''(p_f,+) Q u''(p_i,+)
$$
    we have
$$
\displaystyle
Z = {\cal P}(r,l) \cdot
 \sqrt{ (p_i)_{\bot}^{\ast}  \over  \left| (p_i)_{\bot} \right| }
 \cdot  \sqrt{ (p_f)_{\bot}  \over  \left| (p_f)_{\bot} \right| }
\;\; ,
$$
    where
$
r^{\mu} = (1,0,0,0)
\;\; , \;\;\;
l^{\mu} = (0,0,0,1)
\;\; .
$

\item[3.]
    In the paper \cite{r6.3-1} it is proposed to use the  bispinor
for a massive particle in the form:
$$
\displaystyle
u(p,n)= { {\cal P}(p,n) \over \sqrt{ (pq)-m(nq) } }
{1 \over \sqrt{ q_{+} } }
 \left( \begin{array}{c} 0 \\ 0 \\ q_{+} \\ q_{\bot}
 \end{array} \right)
\;\; ,
$$
    where \\
$
\displaystyle
q^{\mu} = (q_0=1+\left| \alpha \right|^2, q_x=2Re \, \alpha ,
           q_y=2I\!m \, \alpha , q_z=1-\left| \alpha \right|^2 )
$
    is a massless 4-vector, \\
    $\alpha$ is an arbitrary complex number and
$
\displaystyle
{\cal P}(p,n)
$
    has the form (\ref{e6.1}).

    In this paper it was shown that
$$
\displaystyle
\bar{u}(p_f,n_f) Q u(p_i,n_i) =
 {Tr \left[ Q {\cal P}(p_i,n_i) {\cal P}_{+}(q)
           {\cal P}(p_f,n_f) \right] \over
\sqrt{ Tr \left[ {\cal P}(p_i,n_i) {\cal P}_{+}(q) \right]
       Tr \left[ {\cal P}(p_f,n_f) {\cal P}_{+}(q) \right] } }
\;\; ,
$$
    where
$
\displaystyle
{\cal P}_{+}(q) = { 1 \over 2 } ( 1 + \gamma_5 ) \hat{q}
\;\; .
$
\end{description}

    So  the  methods  of  the  calculation of amplitudes discussed
above  in  this  Section  are  described  by  the  general  scheme
(\ref{e2.2}).  Moreover, most of them are the special cases of the
covariant method proposed in Section~{\bf 4}.  However the defects
of  these   methods  are   complicated  calculations,   bulky  and
noncovariant form of their results.

    In conclusion let  us consider two  methods based on  bispinor
transformation.  These methods  are similar to the  ones discussed
above,  but  do  not  use  the  explicit form of the bispinors and
projection operators.

\begin{description}
\item[4.]

    In the  paper \cite{r6.4-1}  it is  proposed to  transform the
bispinor for a massive particle as follows
\begin{equation}
\displaystyle
u(p,n)
= { {\cal P}(p,n) \over \sqrt{ (pq) + m(nq) } }  u_{-}(q)
= { {\cal P}(p,n) \over
\sqrt{ Tr \left[ {\cal P}(p,n) {\cal P}_{-}(q) \right] } }
     u_{-}(q)
\;\; .
\label{e6.37}
\end{equation}
    In  fact,  the  expression  for  the  bispinor  is  given   by
(\ref{e6.37}) up to a phase  factor.  Really, the right-hand  side
of Eq.  (\ref{e6.37}) can be written as
$$
\displaystyle
{ {\cal P}(p,n) \over
\sqrt{ Tr \left[ {\cal P}(p,n) {\cal P}_{-}(q) \right] } }
     u_{-}(q)
= { u(p,n) \bar{u}(p,n) \over \left| \bar{u}(p,n) u_{-}(q) \right| }
     u_{-}(q)
= { \bar{u}(p,n) u_{-}(q) \over
\left| \bar{u}(p,n) u_{-}(q) \right| }  u(p,n)
\;\; .
$$
    Further
\begin{equation}
\begin{array}{l} \displaystyle
\displaystyle
u(p,n)
= { (\hat{p} + m ) ( 1 + \gamma_5 \hat{n} ) \over
                        2 \sqrt{ (pq) + m(nq) } } u_{-}(q)
= { ( \hat{p} + m ) \hat{q} ( \hat{p} + m )
( 1 + \gamma_5 \hat{n} ) \over
              4 (pq)  \sqrt{ (pq) + m(nq) } } u_{-}(q)
        \\       \\     \displaystyle
= { ( \hat{p} + m ) \left[ { 1 \over 2 } ( 1 - \gamma_5 ) \hat{q}
 + \hat{l} { 1 \over 2 } ( 1 - \gamma_5 ) \hat{q} \hat{l} \right]
{\cal P}(p,n) \over
              4 (pq) \sqrt{ (pq) + m(nq) } } u_{-}(q)
        \\       \\     \displaystyle
= { ( \hat{p} + m ) \left[ {\cal P}_{-}(q)
 + \hat{l} {\cal P}_{-}(q) \hat{l} \right]
{\cal P}(p,n) \over
              4 (pq)  \sqrt{ (pq) + m(nq) } }  u_{-}(q)
        \\       \\     \displaystyle
= { ( \hat{p} + m )
\left\{ Tr \left[ {\cal P}_{-}(q) {\cal P}(p,n) \right]
 + \hat{l} Tr \left[ {\cal P}_{-}(q) \hat{l} {\cal P}(p,n) \right]
 \right\}  \over  4 (pq)  \sqrt{ (pq) + m(nq) } }  u_{-}(q)
        \\       \\     \displaystyle
= { ( \hat{p} + m )
\left\{  \left[ (pq) + m(nq)  \right]
 +  \left[ (pl)(nq) - (pq)(nl) -
 {\it i} \varepsilon(p,n,q,l)  \right] \hat{l} \right\}
 \over  4 (pq)  \sqrt{ (pq) + m(nq) } }  u_{-}(q)
\;\; ,
\end{array}
\label{e6.38}
\end{equation}
    where  $q$  is  an  arbitrary  massless  4-vector;  $l$  is an
arbitrary 4-vector such that \\
$
\displaystyle
l^2 = -1
\; , \;\;
$
$
\displaystyle
ql = 0
\;
$
    [to obtain (\ref{e6.38}), we used a variation of the  identity
(\ref{e4.2})
$$
\displaystyle
{\cal P}_{\pm}(q) A u_{\pm}(q)
= Tr [ {\cal P}_{\pm}(q) A ] \cdot u_{\pm}(q)
$$
    where $A$ is an arbitrary matrix operator].

    As  it  follows   from  (\ref{e6.37}),  the   above  presented
transformation  of  the  bispinor   leads  to  the  method  of the
calculation of the amplitudes  given by  a special case of formula
(\ref{e5.8}).
\item[5.]
    In the  paper \cite{r6.5-1}  it is  proposed to  transform the
bispinor for a massive particle as follows
\begin{equation}
\displaystyle
u(p,n)
= { \bar{u}_{+}(q^p) u_{-}(q^m) \over m }  u_{+}(q^p)
   +  u_{-}(q^m)
\;\; ,
\label{e6.39}
\end{equation}
    where
$$
\displaystyle
q^p = { 1 \over 2} (p + mn)
\;\; , \;\;
q^m = { 1 \over 2} (p - mn)
\;\; , \;\;
(q^p)^2 =(q^m)^2 = 0
\;\; , \;\;
(q^p q^m) = { m^2 \over 2}
\;\; .
$$
    This transformation can be  obtained by the method  similar to
the one discussed in previous subsection.  Really,
$$
\begin{array}{l} \displaystyle
u(p,n)
={ {\cal P}(p,n) \over
\sqrt{ Tr \left[ {\cal P}(p,n) {\cal P}_{-}(q^m) \right] } }
     u_{-}(q^m)
= { { \hat{q} }^p + m \over m} u_{-}(q^m)
= { { 1 \over 2 } ( 1 + \gamma_5 ){ \hat{q} }^p + m \over m }
u_{-}(q^m)
         \\      \\     \displaystyle
= \left[ { u_{+}(q^p) \bar{u}_{+}(q^p) \over m } + 1 \right]
u_{-}(q^m)
=  { \bar{u}_{+}(q^p) u_{-}(q^m) \over m } u_{+}(q^p)
   +  u_{-}(q^m)
\;\; .
\end{array}
$$
    Because of this, the transformation of the bispinor considered
leads to the method of the calculation of the amplitudes given  by
a special case of formula (\ref{e5.8}).

    However, to calculate
$
\displaystyle
\bar{u}_{+}(q^p) u_{-}(q^m)
$
    the authors of \cite{r6.5-1} use the method which was proposed
in the  papers \cite{r2.10a}~--~\cite{r2.10b}.   (This  method was
discussed in subsection 10 of  Section~{\bf 2}.)  In this  case we
have up to a phase factor [see~(\ref{e5.3})]
$$
\begin {array}{c} \displaystyle
\bar{u}_{+}(q^p) u_{-}(q^m)
\simeq { Tr [ ( 1 - \gamma_5 ) \hat{q}^m
 \hat{l} \hat{q} \hat{q}^p ] \over
4 \sqrt { ( q q^p ) ( q q^m ) } }
                        \\[0.5cm] \displaystyle
  = [ (q^p)_y + {\it i} (q^p)_z]
  \sqrt{ { (q^m)_0 - (q^m)_x \over (q^p)_0 - (q^p)_x } }
     - [ (q^m)_y + {\it i} (q^m)_z]
  \sqrt{ { (q^p)_0 - (q^p)_x \over (q^m)_0 - (q^m)_x } }
\end {array}
$$
    [where
$
\displaystyle
l^{\mu} = (0,0,1,0) \; , \;\; q^{\mu} = (1,1,0,0)
$~].

    As a result, the first term in (\ref{e6.39}) obtains the
phase factor [see~(\ref{e3.3})]
$$
\displaystyle
 { \bar{u}_{-}(q^m) \hat{l} \hat{q} u_{+}(q^p) \over
 | \bar{u}_{-}(q^m) \hat{l} \hat{q} u_{+}(q^p) | }
$$
    and  the  bispinor  under  consideration  does not satisfy the
Dirac equation.

    Note that  a special  case of  the transformation  considered,
namely the situation when a  4-vector, specifying the axis of  the
spin projections, has form
$$
\displaystyle
n = \lambda \left[ { 1 \over m } p
                     - { m \over (pq^m) } q^m \right]
\;\; , \;\;
\lambda = \pm 1
$$
    was proposed in \cite{r2.10b}.
\end{description}
{\bf 6. Incorrect methods of calculation of the amplitudes}

    In \cite{rm.1-1}  for deriving  of expressions  for amplitudes
incorrect   identity    {\it   (15}),    writting   without    any
substantiation, is used
$$
\displaystyle
{\vec c}{\vec c}'
= - (s s') + { (p s') (p' s) \over (p p') + m m' }
\hspace{2cm}   {\it (15)}
$$
    where
$
\displaystyle
s_0 = { {\vec p}{\vec c} \over m}
\, ,   \;
{\vec s} = {\vec c} + { s_0 \over p_0 + m } {\vec p}
\, ; \;\;\;
s'_0 = { {\vec p}'{\vec c}' \over m'}
\, ,   \;
{\vec s}' = {\vec c}' + { s'_0 \over p'_0 + m' } {\vec p}'
\, .
$
\vspace{5mm}

    Authors of \cite{rm.1-2} propose to use bispinors for  massive
Dirac particles in the form
$$
\displaystyle
u_{+} (p) =
\left[
\begin{array}{l}
\displaystyle
 { {\it i} m \over \sqrt{ 2 (pq) q_{-} } }
\left(
\begin{array}{l}
        - q_{-}  \\   q_{\bot}
\end{array}
\right)     \\
\displaystyle
 { 1 \over \sqrt{ k_{-} } }
\left(
\begin{array}{l}
         k_{\bot}^{\ast} \\  k_{-}
\end{array}
\right)
 \end{array}
\right]
\;\; ,
$$
$$
\displaystyle
u_{-} (p) =
\left[
\begin{array}{l}
\displaystyle
 { {\it i} \over \sqrt{ k_{-} } }
\left(
\begin{array}{l}
        k_{-}    \\   - k_{\bot}
\end{array}
\right)     \\
\displaystyle
 { m \over \sqrt{ 2 (pq) q_{-} } }
\left(
\begin{array}{l}
         q_{\bot}^{\ast} \\  q_{-}
\end{array}
\right)
 \end{array}
\right]
$$
    where
$$
\displaystyle
p = k + { m^2 \over 2 (pq) } q
\, , \;
p^2 = m^2
\, , \;
q^2 = 0
\, ,
$$
$$
\displaystyle
k = p - { m^2 \over 2 (pq) } q
\, , \;
k^2 = 0
\, .
$$
    It  is  stated,  that  $q$  is  an  arbitrary massless vector.
However from elementary requirement
$$
\displaystyle
  \bar{u}_{+}(p) {\gamma}_5 u_{+}(p)
= \bar{u}_{-}(p) {\gamma}_5 u_{-}(p) = 0
$$
    we have
$$
\displaystyle
  { q_x \over q_0 - q_z } = { p_x \over p_0 - p_z }
$$
    i.e. vector $q$  is not arbitrary.   Besides, 4-vector,  which
determines the  axes  of  the  spin projections, is not defined in
this paper.
\vspace{5mm}

    In paper \cite{rm.1-3} there are mistakes in  phase factors of
coefficients
$
\displaystyle
C_{ {\lambda} {\lambda}' }
$,
    given by  equations  ${\it (10)}$  therein.  As  a consequence
method gives incorrect results.  For example, using this method we
have
$$
\begin {array}{r} \displaystyle
[ {\bar u} ( p , + )  {\gamma}_5  u ( p', + ) ]
[ {\bar u} ( p , + )  {\gamma}_5  u ( p', + ) ]^{\ast}
= { | {\vec p} | | {\vec p}'| + ( {\vec p} {\vec p}')
 \over
| {\vec p} | | {\vec p}'|
( p p' + m m')^2 \sqrt { p_x^2 + p_y^2 } {\vec p}^2 }
\cdot
   \\[0.5cm] \displaystyle
\Big\{
( p_0 p_0' - | {\vec p} | | {\vec p}'| + m m' )
\sqrt{ p_x^2 + p_y^2 }
\left[
 p_0 ( {\vec p} {\vec p}') - p_0' {\vec p}^2
\right]^2
   \\[0.5cm] \displaystyle
+
( p_0 p_0' + | {\vec p} | | {\vec p}'| + m m' )
\sqrt{ p_x^2 + p_y^2 }
\left[
| {\vec p} | | {\vec p}'| - ( {\vec p} {\vec p}')
\right]^2 m^2
   \\[0.5cm] \displaystyle
-
2 ( m p'_0 + m' p_0 ) p_x
\left[
 p_0 ( {\vec p} {\vec p}') - p_0' {\vec p}^2
\right]
\left[
| {\vec p} | | {\vec p}'| - ( {\vec p} {\vec p}')
\right] m
\Big\}
\end {array}
$$
    However, calculation by the classical  method  gives   another
result:
$$
\displaystyle
| {\bar u} ( p , + )  {\gamma}_5  u ( p', + ) |^2
= - { [ m m' - p_0 p_0' + | {\vec p} | | {\vec p}'| ]
    [ ( {\vec p} {\vec p}') + | {\vec p} | | {\vec p}'| ]
\over
 | {\vec p} | | {\vec p}'| }
\, .
$$



    \section {The  computer programs  for the  calculation of  the
amplitudes}

    A  number  of  computer  programs  for  the calculation of the
amplitudes   of   processes   with   polarized   Dirac   particles
(antiparticles) has been developed in the last few years.  However
all of them have these or those disadvantages and limitations.  In
particular, they do  not allow us  to calculate the  amplitudes of
processes involving massive  Dirac particles (antiparticles)  with
arbitrary polarization  states.   Let us  consider these  programs
briefly.

\begin{description}
\item[1.]

    The  program  CHANEL  (see~\cite{r7.1-1})  is  a  part  of the
computer system GRACE (see~\cite{r7.1-2}).  This program allows us
to  calculate  the  amplitudes  of  processes  with  massive Dirac
particles (antiparticles), the polarization state of which is  the
helicity,  i.e. the  4-vectors  specifying   the axes of  the spin
projection are
$$
\displaystyle
n^{\mu}(p)={ \lambda \over m} \left( \left| \vec{p} \right|,
  {p_0 \over \left| \vec{p} \right|} \vec{p} \right)
\;\; , \;\;\;
\lambda = \pm 1
$$
    [where
$
\displaystyle
p^{\mu} = ( p_0 , \vec{p} )
$
    is   the    4-momentum   of    the   corresponding    particle
(antiparticle)].

    The program uses the method of calculation, which is analogous
to the one considered in subsection 5 of Section~{\bf 6}:
\begin{description}
\item[(a)]
$$
\displaystyle
\lambda = +1
$$
$$
\displaystyle
( q^p_{+1} )^{\mu}
= { p_0 + | \vec{p} | \over 2 | \vec{p} | }
       \left( | \vec{p} |, \vec{p}  \right)
\;\; , \;\;
( q^m_{+1} )^{\mu}
= { p_0 - | \vec{p} | \over 2 | \vec{p} | }
       \left( | \vec{p} |, - \vec{p}  \right)
\;\; ,
$$
\begin{equation}
\begin{array}{l} \displaystyle
u( p , \lambda = +1 )
={ {\cal P}( p , \lambda = +1 ) \over
\sqrt{ Tr \left[ {\cal P}( p , \lambda = +1 )
{\cal P}_{+}( q^p_{+1} ) \right] } }  u_{+}( q^p_{+1} )
= { { \hat{q} }^m_{+1} + m \over m} u_{+}( q^p_{+1} )
         \\      \\     \displaystyle
= { {1 \over 2} (1 - \gamma_5 ) { \hat{q} }^m_{+1} + m \over m }
u_{+}( q^p_{+1} )
= \left[ { u_{-}( q^m_{+1} ) \bar{u}_{-}( q^m_{+1} ) \over m } + 1
\right] u_{+}( q^p_{+1} )
         \\      \\     \displaystyle
=  { \bar{u}_{-}( q^m_{+1} ) u_{+}( q^p_{+1} ) \over m }
   u_{-}( q^m_{+1} ) +  u_{+}( q^p_{+1} )
\;\; .
\end{array}
\label{e7.1-1}
\end{equation}
\item[(b)]
$$
\displaystyle
\lambda = -1
$$
$$
\displaystyle
( q^p_{-1} )^{\mu}
= { p_0 - | \vec{p} | \over 2 | \vec{p} | }
       \left( | \vec{p} |, - \vec{p}  \right)
\;\; , \;\;
( q^m_{-1} )^{\mu}
= { p_0 + | \vec{p} | \over 2 | \vec{p} | }
       \left( | \vec{p} |, \vec{p}  \right)
\;\; ,
$$
\begin{equation}
\begin{array}{l} \displaystyle
u( p , \lambda = -1 )
={ {\cal P}( p , \lambda = -1 ) \over
\sqrt{ Tr \left[ {\cal P}( p , \lambda = -1 )
{\cal P}_{-}( q^m_{-1} ) \right] } }  u_{-}( q^m_{-1} )
= { { \hat{q} }^p_{-1} + m \over m} u_{-}( q^m_{-1} )
         \\      \\     \displaystyle
= { {1 \over 2} ( 1 + \gamma_5 ) { \hat{q} }^p_{-1} + m \over m }
u_{-}( q^m_{-1} )
= \left[ { u_{+}( q^p_{-1} ) \bar{u}_{+}( q^p_{-1} ) \over m } + 1
\right] u_{-}( q^m_{-1} )
         \\      \\     \displaystyle
=  { \bar{u}_{+}( q^p_{-1} ) u_{-}( q^m_{-1} ) \over m }
   u_{+}( q^p_{-1} ) +  u_{-}( q^m_{-1} )
\;\; .
\end{array}
\label{e7.1-2}
\end{equation}
\end{description}
    Introducing the notation
$$
\displaystyle
q_1 =  q^p_{+1} = q^m_{-1}
\;\; , \;\;
q_2 =  q^m_{+1} = q^p_{-1}
$$
    and combining (\ref{e7.1-1}) and (\ref{e7.1-2}) we obtain
\begin{equation}
\displaystyle
u( p , \lambda )
=  { \bar{u}_{ - \lambda }( q_2 ) u_{\lambda}( q_1 ) \over m }
   u_{ - \lambda }( q_2 ) +  u_{\lambda}( q_1 )
\;\; .
\label{e7.1-3}
\end{equation}
    Further we have [up to phase factor -- see~also~(\ref{e5.3})]
$$
\begin{array}{l} \displaystyle
\bar{u}_{ - \lambda }( q_2 ) u_{\lambda}( q_1 )
\simeq
\lambda { Tr \left[ \hat{q}_1 \hat{q} \hat{l} \hat{q}_2
                  ( 1 + \lambda \gamma_5 ) \right]  \over
         4 \sqrt{ ( q q_1 ) ( q q_2 ) } }
= \lambda m { Tr \left[ \hat{q} \hat{l} \hat{r} \hat{l}_p
                  ( 1 - \lambda \gamma_5 ) \right]  \over
          4 \sqrt{ ( q r )^2 - ( q l_p )^2 } }
               \\[0.5cm]    \displaystyle
= m { \lambda p_y - {\it i} p_z  \over
    \sqrt{ {p_y}^2 + {p_z}^2 } }
\; .
\end{array}
$$
    Here
$$
\displaystyle
 q_1 = { p_0 + | \vec{p} | \over 2 } ( r + l_p )
\; , \;\;
 q_2 = { p_0 - | \vec{p} | \over 2 } ( r - l_p )
\; ,
$$
$$
\displaystyle
 r^{\mu} = ( 1, 0, 0, 0 )
\; , \;\;
 {l_p}^{\mu} = ( 0, { \vec{p} \over | \vec{p} | } )
\; ,
$$
    $q$, $l$ are 4-vectors such that
$$
\displaystyle
q^2 = 0  \; , \;\; l^2 = -1   \; , \;\; (ql) = 0
$$
    [in this case one chooses
$
\displaystyle
q^{\mu} = (1,1,0,0) \; , \;\; l^{\mu} = (0,0,1,0)
$].

    However in this case the first term in (\ref{e7.1-3}) obtains
the phase factor
$$
\displaystyle
  \lambda
 { \bar{u}_{\lambda}(q_1) \hat{q} \hat{l} u_{-\lambda}(q_2)
 \over
 | \bar{u}_{\lambda}(q_1) \hat{q} \hat{l} u_{-\lambda}(q_2) | }
\; .
$$
    As a result, the bispinor (\ref{e7.1-3}) does not satisfy  the
Dirac equation.

\item[2.]
    The program COMPUTE (see~\cite{r7.2-1}) use the method of  the
calculation  of  the  amplitudes  of  processes with massive Dirac
particles   (antiparticles)   proposed   in   \cite{r6.1.5-1}  and
considered in subsection 1.5 of Section~{\bf 6}.

    Remind that in this  case the polarization state  of particles
(antiparticles) is defined by the vectors
$$
\displaystyle
n^{\mu} = { \lambda \over m } ( p_0 - { m^2 \over p_0 + p_z },
p_x , p_y , p_z + { m^2 \over p_0 + p_z } )
\, , \;
\lambda = \pm 1
$$
    [where
$
\displaystyle
p^{\mu} = ( p_0 , p_x , p_y , p_z )
$
    is   the    4-momentum   of    the   corresponding    particle
(antiparticle)].

\item[3.]
    The  program  HELAS  (see~\cite{r7.3-1})  is  a  part  of  the
computer system MadGraph  (see~\cite{r7.3-2}).  It  calculates the
amplitudes   of   processes    with   massive   Dirac    particles
(antiparticles), the polarization state  of which is the  helicity
by  the  method  proposed  in  \cite{r6.1.2-1}.    This  method is
considered in subsection 1.2 of Section~{\bf 6}.

\item[4.]
    There are two variants for the program HIP.  The first variant
(see~\cite{r7.4-1}) calculates  the amplitudes  of processes  with
massless Dirac particles (antiparticles) by the method proposed in
\cite{r2.10b} and considered in subsection 10 of Section~{\bf 2}.

    The second one  (see~\cite{r7.4-2}) calculates the  amplitudes
of  processes  with  massive  Dirac particles (antiparticles), the
polarization  state  of  which  is  the  helicity,  by  the method
proposed in  \cite{r6.1.2-1} and  considered in  subsection 1.2 of
Section~{\bf 6}.

\item[5.]
    The   program   SPINORP   (see~\cite{r7.5-1})  calculates  the
amplitudes   of   processes   with   massless   Dirac    particles
(antiparticles)  by  the  method  proposed  in \cite{r6.1.3-2} and
considered in subsection 1.3 of Section~{\bf 6}.

\item[6.]
    The  program  WPHAST  1.0  (see~\cite{r7.6-1})  calculates the
amplitudes  of   processes  involving   massive  Dirac   particles
(antiparticles)  by  the  method  proposed  in \cite{r2.5.2-1} and
considered in subsections 5.2 and 10 of Section~{\bf 2}.

    Recall that in this case the polarization state of particles
(antiparticles) is defined by the vectors
$$
\displaystyle
n = \lambda \left[ { 1 \over m } p
                     - { m \over (pq) } q \right]
\;\; , \;\;
\lambda = \pm 1
$$
    [where $p$  is the  4-momentum of  the corresponding  particle
(antiparticle) and $q$ is an arbitrary 4-vector such that
$
\displaystyle
q^2 = 0
$
    and in numerical calculations one chooses
$
\displaystyle
q^{\mu} = (1,1,0,0)
$
].

\item[7.]
    In~\cite{r7.7-1}  there  was  proposed  the  program  for  the
calculation  of  the  amplitudes  of  processes with massive Dirac
particles   (antiparticles)    by   the    method   proposed    in
\cite{r6.1.4-1} and considered  in subsection 1.4  of Section~{\bf
6}.

    Recall that in this  case the polarization state  of particles
(antiparticles) is defined by the vectors
$$
\displaystyle
n^{\mu} = { \lambda \over m } (p_z, { p_x p_z \over p_0 + m },
{ p_y p_z \over p_0 + m }, m + { p_z^2  \over p_0 + m } )
\, , \;
\lambda = \pm 1
$$
[where
$
\displaystyle
p^{\mu} = ( p_0 , p_x , p_y , p_z )
$
    is   the    4-momentum   of    the   corresponding    particle
(antiparticle)].
\end{description}

    However, in addition to programs described above some  authors
use the computer  algebra systems to  calculate the amplitudes  of
processes  for  concrete  problems.    For  example, the author of
\cite{r6.1.2-3} uses the system MAPLE for  the calculation of  the
amplitudes  of   processes  involving   massive  Dirac   particles
(antiparticles), the polarization state of which is the  helicity,
by the method considered in subsection 1.2 of Section~{\bf 6}.

    Thus,  at  present  there  are  no  computer  programs for the
calculation of the amplitudes of processes involving massive Dirac
particles  (antiparticles)   which  have   arbitrary  polarization
states.  The programs of this sort may be created with help of the
formulae presented in Section~{\bf 5}.



\end {document}